\documentclass[reprint,superscriptaddress,aps,prb]{revtex4-2}

\usepackage{amsmath, amssymb, siunitx, graphicx, braket, pgf, hyperref}
\newcommand{\bcal}[1]{\boldsymbol{\cal #1}}
\newcommand{\heli}{\varphi}
\newcommand{\Opheli}{\hat{\varphi}}
\newcommand{\vOp}[1]{\hat{\boldsymbol{#1}}}
\newcommand{\Ham}[1]{\hat{H}_{\rm #1}}

\newcommand{\Sz}{\hat{\cal S}_z}
\newcommand{\Apar}{\bar{\boldsymbol{\Phi}}}

\newcommand{\collcoor}{c_{\heli}}
\newcommand{\vs}{\xi}
\newcommand{\vp}{\zeta}

\mathchardef\shyph="2D

\begin{document}

\title{Optical signatures of quantum skyrmions}

\author{Sanchar Sharma}
\email{sanchar.sharma@phys.ens.fr}
\affiliation{Laboratoire de Physique de l’\'{E}cole Normale Sup\'{e}rieure, ENS, Universit\'{e} PSL, CNRS, Sorbonne Universit\'{e}, Universit\'{e} de Paris, F-75005 Paris, France}

\author{Christina Psaroudaki}
\email{christina.psaroudaki@phys.ens.fr}
\affiliation{Laboratoire de Physique de l’\'{E}cole Normale Sup\'{e}rieure, ENS, Universit\'{e} PSL, CNRS, Sorbonne Universit\'{e}, Universit\'{e} de Paris, F-75005 Paris, France}

\date{\today}

\begin{abstract}
Magnets have recently emerged as promising candidates for quantum computing, particularly using topologically-protected nanoscale spin textures. While the quantum dynamics of such spin textures has been theoretically studied, direct experimental evidence of their non-classical behavior remains an open challenge. To address this, we propose to employ Brillouin light scattering (BLS) as a method to probe the quantum nature of skyrmions in frustrated magnets. We show that, for a specific geometry, classical skyrmions produce symmetric sidebands in the BLS spectrum, whereas quantum skyrmions exhibit a distinct asymmetry arising from vacuum fluctuations of their rotation. By studying the photon-skyrmion interaction, we calculate the BLS spectrum using a quantum master equation and show that sideband asymmetry serves as a robust witness of energy level quantization. We find that this asymmetry is pronounced at low temperatures, and can be controlled by input laser power. These findings establish a concrete protocol for the optical detection of non-classical features in spin textures, paving the way for exploring their role in quantum applications.
\end{abstract}

\maketitle

Ferromagnets are well-established in classical applications due to their low losses and scalability~\cite{Roadmap}, and have recently attracted interest for quantum computing, with several experimental~\cite{QMag_Tabuchi, QMag_Quirion, Xu_MagFock, Xu_Bell} and theoretical~\cite{MagEnt_Mehrdad, MagHer_Victor, CatSt_Mine, ArbitStGen_Mine, QMag_Marios, Coherence_Mehrdad, QTom_Mag, Martijn_QGates} advances. Most studies focus on the uniform magnetic mode, but non-uniform textures such as skyrmions and domain walls~\cite{Skyr_QDyn_PRX, SkyrQBrown_PRB, QSkyr_Lohani, SkyrQub_PRL, SkyrQub_PRA, DW_QTrans, SkyrHel_UnivQC, SkyrQub_Coll} also offer potential for quantum applications. Magnetic skyrmions are topologically protected defects, whose nanoscale size and low energy cost make them attractive for data storage and transport applications~\cite{Fert_SkyrRev, Bogdanov_SkyrRev}. In frustrated magnets, skyrmions have been proposed~\cite{FrustSkyr_Tsuyoshi, FrustSkyr_Leonov} and experimentally shown~\cite{FrustSkyr_Kurumaji} to be exceptionally small, making them promising candidates for exhibiting quantum behavior. Theoretically, quantum skyrmions were explored~\cite{SkyrQub_Coll, SkyrQuant_PRB}, including proposals of universal skyrmion-based qubits~\cite{SkyrQub_PRL}, but their non-classical behaviour has not yet been demonstrated experimentally.

The small size of skyrmions makes direct observation of their quantum states challenging due to weak signals. However, quantum witnesses can still exist in steady-state measurements, as demonstrated in trapped ions~\cite{Diedrich_IonCool, Eschner_IonCool} and mechanical systems~\cite{Clerk_QNoise_Rev, CavOMech_Aspelmeyer}. In the latter, Brillouin light scattering (BLS)—the inelastic scattering of photons via material excitations—reveals quantum behavior through sideband asymmetry (SA) in the photon spectrum. While classical mechanical systems produce symmetric sidebands, experiments have observed SA as a clear hallmark of quantum effects~\cite{SA_Safavi, SA_Weinstein}. 

\begin{figure}
    \centering
    \includegraphics[width=\linewidth]{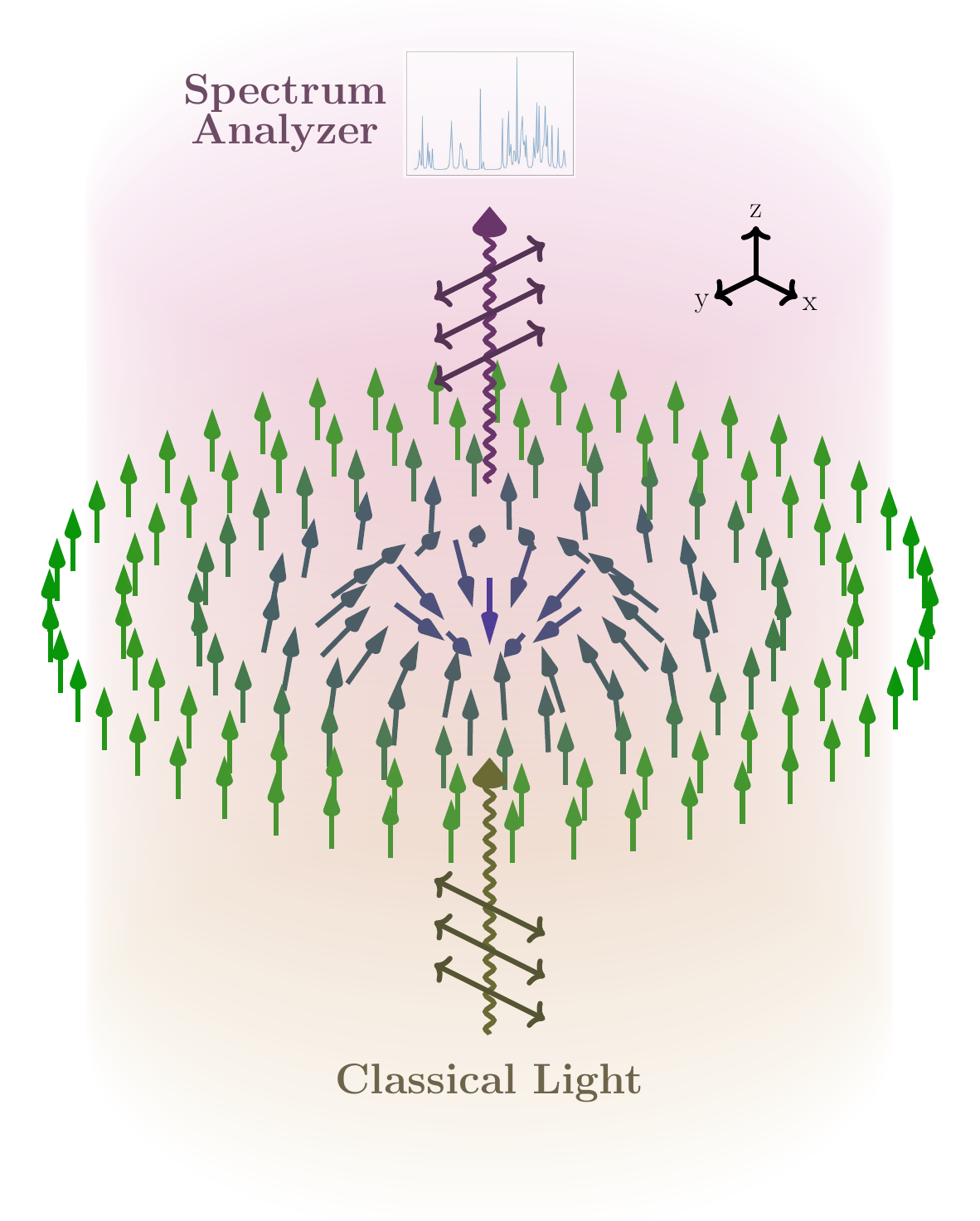}    
    \caption{A large laser input (classical light) impinges on a magnet hosting a skyrmion. The FM ground state is along $+z$. The spectrum of the scattered light, polarized perpendicular to the input polarization, is measured.}
    \label{fig:Setup}
\end{figure}

BLS is a well-established classical probe of the magnetization~\cite{BLS_Mag_Rev, MicroBLS_Rev}, which has allowed for detecting and mapping spin-wave eigenmodes in several geometries. Extending BLS techniques to quantum measurements is highly appealing owing to its high spatio-temporal resolution ($<\si{\micro\meter}$ and $<\si{\nano\second}$), and absence of passive dissipation, as the spin-photon coupling can be controlled via an external laser.
Recently, BLS was theoretically shown to be promising for reconstructing the density matrix of the uniform magnetization~\cite{QTom_Mag}. 
Here, we show that BLS can reveal quantum features of magnetic skyrmions.

Using SA as a witness for non-classicality of a magnetic system, in analogy with mechanics, presents a potential challenge: BLS via spins typically exhibits SA even under classical conditions~\cite{Wettling_SA, BLS_Mag_Rev, OptMag_Osada, OptMag_Zhang, OptMag_James}. 
This asymmetry arises from the interference of different microscopic spin-photon interactions, specifically the circular and linear birefringences ~\cite{Wettling_SA, BLS_WGM_Mine}.
Thus, as a first step of our theory, we show below that for the geometry shown in Fig.~\ref{fig:Setup}, classical SA vanishes.
On the other hand, quantum SA arises from the vacuum fluctuations, and therefore it is expected to be large, especially at low temperatures where thermal noise is small. 
To show this, we derive the optical BLS spectrum from a skyrmion in a frustrated magnet and the predicted SA is a witness of the quantum nature of skyrmions.

\noindent \textbf{Skyrmion Quantized States:} 
We briefly review the quantization of skyrmion coordinates used to characterize the quantum states of skyrmion rotation, following \cite{SkyrQub_PRL, SkyrQub_PRA}. 
We have provided the detailed derivations for the results discussed below in the Appendix.
We consider spins $\{\vOp{S}_{
\alpha}\}$ of magnitude $\bar{S}$ on a two-dimensional (2D) triangular lattice, governed by the Hamiltonian,
\begin{equation}
    \hat{H}_{\rm mag} = \frac{1}{2} \sum_{\alpha\beta} J_{\alpha\beta} \vOp{S}_{\alpha} \cdot \vOp{S}_{\beta} - \frac{K}{2} \sum_{\alpha} \hat{S}_{\alpha z}^2 - B\sum_{\alpha} \hat{S}_{\alpha z},\label{Eq:Hamiltonian}
\end{equation}
where the set $\{J_{\alpha\beta}\}$ are the exchange couplings, $K$ is the strength of easy-axis anisotropy, and $B$ is proportional to an external magnetic field. 
We consider ferromagnetic (FM) nearest-neighbor ($J_1<0$) and antiferromagnetic next-nearest-neighbor ($J_2>0$) exchange interactions.
At high magnetic fields, the ground state is fully polarized along $+z$. A magnetic skyrmion appears as a defect on this background~(Fig.~\ref{fig:Setup}) and remains stable over a broad parameter space~\cite{FrustSkyr_Leonov}.

The generator of global spin rotations,
\begin{equation}
    \hat{\mathcal{S}}_z = \sum_{\alpha} (\bar{S} - \hat{S}_{\alpha z})\,, \label{def:Sz} 
\end{equation}
commutes with $\hat{H}_{\rm mag}$.
The ground state is degenerate under the transformation $e^{-i\hat{\cal S}_z \heli}$, where $\heli$ is the zero mode of skyrmion helicity~\cite{SkyrQuant_PRB}. We consider the path-integral formulation of Eq.~\eqref{Eq:Hamiltonian}, allowing for simple definitions of collective coordinates by replacing spin operators with coherent state variables~\cite{SkyrQub_PRL, SkyrQub_PRA}. We use the spin parametrization $S_{\alpha z} = \Pi_{\alpha}$ and $S_{\alpha x} \pm iS_{\alpha y} = \sqrt{1-\Pi_{\alpha}^2} e^{i\Phi_{\alpha}}$, where $\Phi_{\alpha}$ gives the azimuthal angle of the spins and its conjugate momentum $\Pi_{\alpha}$ quantifies the deviation from the FM state.
We numerically find the stabilized skyrmion state, up to a helicity rotation, denoted by $\{\Pi_{0\alpha},\Phi_{0\alpha}\}$ as a minima of the path-integral Hamiltonian, using the code in \cite{Hullahalli2025}.
For a spin located at $(\rho_{\alpha}, \phi_{\alpha})$ in 2D polar coordinates, we find that typically $\Pi_{0\alpha}$ depends only on $\rho_{\alpha}$ and $\Phi_{0\alpha} \approx \phi_{\alpha}$. 

The skyrmion helicity is defined as an argument of a weighted sum
\begin{equation}
    \heli= \arg\left[ \sum_{\alpha} \sqrt{1-\Pi_{0\alpha}} \sqrt{1-\Pi_{\alpha}} e^{i(\Phi_{\alpha} - \Phi_{0\alpha})} \right]\,,\label{def:heli}
\end{equation}
which ensures that the phase outside the skyrmion, where $1-\Pi_{0\alpha} \approx 0 $, does not contribute to the helicity.  We explicitly show $\{ \varphi,\mathcal{S}_z \}=1$, verifying the two variables are canonically conjugate. Due to azimuthal symmetry, the path-integral Hamiltonian does not depend on $\heli$.
In terms of $\mathcal{S}_z$, it reduces to
\begin{equation}
    H_{\rm skyr}^{\rm path-int} = L(\mathcal{S}_z - \mathcal{S}_{z0})^2 + G(\mathcal{S}_z - \mathcal{S}_{z0})^3, \label{Ham_Sz_path}
\end{equation}
where $\{L,G\}$ are Taylor coefficients, and we ignored higher order terms in $\mathcal{S}_z$ and non-skyrmionic magnon excitations. 
Here, $\mathcal{S}_{z0} = \bar{S}\sum_{\alpha} (1 - \Pi_{0\alpha}) $ and the term linear in $\mathcal{S}_z - \mathcal{S}_{z0}$ vanishes because we are expanding around the local minima $\{\Pi_{0\alpha}, \Phi_{0\alpha}\}$. By the usual correspondence between path integrals and quantum Hamiltonians, we promote helicity to an operator, $\Opheli$, satisfying the canonical commutation relations $[\Opheli,\hat{\cal S}_z]=i$, and replace $H_{\rm skyr}^{\rm path-int} \rightarrow \hat{H}_{\rm skyr}$ obtained by replacing $\mathcal{S}_z \rightarrow \hat{\mathcal{S}}_z$ in Eq.~\eqref{Ham_Sz_path}.

We show below that a finite BLS signal requires breaking the azimuthal symmetry.
This can be achieved by electric fields, magnetic field gradients, or Dzyaloshinskii-Moriya interaction~\cite{SkyrQub_PRL}.
For concreteness, we focus on the electric field, noting that analogous results hold for the other symmetry-breaking mechanisms. It induces a helicity potential $e\cos\Opheli$, where $e$ is proportional to the strength of the electric field, $E$. 
We assume $E \ll \{J_{1,2}, K, B\}$, such that we can ignore its effect on the skyrmion metastable state $\{\Pi_{0\alpha}, \Phi_{0\alpha}\}$.

Thus, our effective Hamiltonian reads,
\begin{equation}
    \Ham{skyr} = L(\hat{\cal S}_z - \mathcal{S}_{z0})^2 + G(\hat{\cal S}_z - \mathcal{S}_{z0})^3 + e \cos\Opheli, \label{Ham_skyr}
\end{equation}
where we find $\{L,G,e\}$ numerically~\cite{Hullahalli2025}.
In the space of helicity wavefunctions $\psi(\heli)$, we can replace $\hat{\cal S}_z = -i\partial_{\varphi}$. 
We can then diagonalize the skyrmion Hamiltonian as $\Ham{skyr} = \hbar \sum_u \omega_u \ket{u}\bra{u}$, with $\hbar \omega_u$ being the energy of mode $u$.
When $e=0$, the eigenmodes are simply $\psi_u(\heli) = e^{iu\heli}$.
For small perturbations, $\cos\Opheli$ hybridizes $\psi_u$ with $\psi_{u\pm 1}$, giving non-zero off-diagonal entries of $\hat{\cal S}_z$, required for optically-induced transitions.

\begin{figure}
    \centering
    \includegraphics[width=\linewidth]{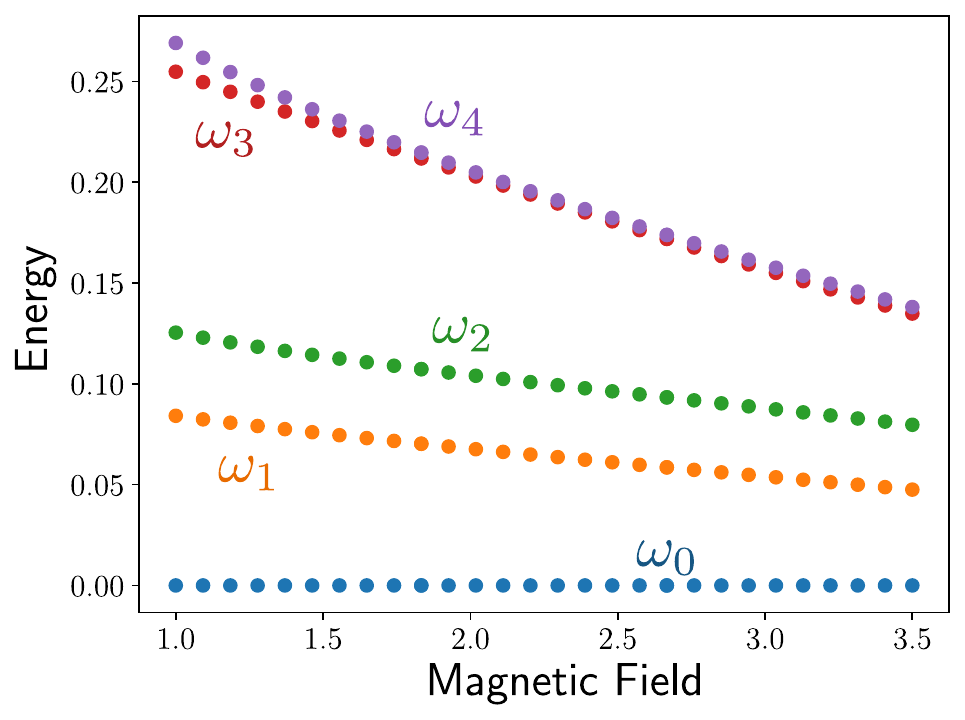}
    \caption{Energy as a function of magnetic field, with reference to $\omega_0$. The microscopic parameters chosen for this plot are $J_1 = 1$, $J_2=0.5$, $\bar{S}=10$, $K=0.7$, and $E=10^{-3}$. See main text for conversion to experimentally accessible values.}
    \label{fig:energy}
\end{figure}

Fig.~\ref{fig:energy} shows the eigenvalues as a function of the applied field, using the parameters specified in the caption. 
In real units, the energy is measured in $J_1 \bar{S}^2 = \SI{15.6}{\meV} = h\times \SI{3.8}{\THz}$, using $A_{\rm ex} = J_1 \bar{S}^2/a $ and $A_{\rm ex} = \SI{5}{\pico\joule\per\meter}$ for the stiffness constant and $a = \SI{0.5}{\nm}$ for the lattice constant. 
The $x$-axis of Fig.~\ref{fig:energy} ranges from $B \in \{0.16, 0.55\} \si{\meV}$, or  $\tilde{B} \in \{1.3, 4.7\} \si{\tesla}$ with $\tilde{B} = B/\gamma\hbar$. 
In these units, the energy range of Fig.~\ref{fig:energy} corresponds to a maximum energy of $\sim h\times\SI{11}{\GHz}$.
Below, we use a similar scale for inverse temperature $\beta$, where $\beta = 1$ corresponds to a temperature of $\SI{1.8}{\kelvin}$.
Note that these ranges strongly depend on $A_{\rm ex}$ and $a$, which would vary across materials. 

\noindent \textbf{Classical sideband symmetry.}  
We now discuss photon scattering by the skyrmion's helicity and show that classical BLS exhibits no SA, i.e. for incident light with frequency $\omega_{\rm in}$, $\bcal{E}_{\rm in} \sim e^{-i\omega_{\rm in}t}$, the output light has a spectrum symmetric around $\omega_{\rm in}$, $|\bcal{E}_{\rm sc}(\omega_{\rm in} + \omega)| = |\bcal{E}_{\rm sc}(\omega_{\rm in} - \omega)|$. A laser incident on a magnet  (Fig.~\ref{fig:Setup}) induces a spin-dependent dipole moment $\boldsymbol{d}(\boldsymbol{r}) = \sum_{\alpha} \delta(\boldsymbol{r} - \boldsymbol{r}_{\alpha}) \boldsymbol{d}_{\alpha}$, where its on-site value is~\cite{Wettling_SA,FleuryLoudon,BLS_WGM_Mine}
\begin{equation}
	\boldsymbol{d}_{\alpha} = if \bcal{E}_{\rm in}(\boldsymbol{r}_{\alpha}) \times \boldsymbol{S}_{\alpha} + g \boldsymbol{S}_{\alpha} (\boldsymbol{S}_{\alpha} \cdot \bcal{E}_{\rm in}(\boldsymbol{r}_{\alpha})), \label{def:dalpha}
\end{equation}
with $f$ and $g$ quantifying circular and linear birefringence respectively.
For concreteness, we focus on an $x$-polarized monochromatic cylindrically symmetric  input, $\bcal{E}_{\rm in} = \mathcal{E}(\rho) \boldsymbol{x} e^{i(k_{\rm in}z - \omega_{\rm in}t)}$ with $\omega_{\rm in} = k_{\rm in}c$ the input frequency and $\mathcal{E}(\rho)$ the transverse dependence of the amplitude.

The far-field radiation moving in the $z$-direction is~\cite{Jackson}, 
\begin{equation}
    \bcal{E}_{\rm sc}(\boldsymbol{r}) \propto \sum_{\alpha} \frac{\boldsymbol{d}_{\alpha}^{\perp}}{|\boldsymbol{r} - \boldsymbol{r}_{\alpha}|} ,
\end{equation}
where $\boldsymbol{d}_{\alpha}^{\perp}$ is the component of the dipole moment perpendicular to $z$.
Note that if either $f$ or $g$ is zero, $|\boldsymbol{d}_{\alpha}(\omega_{\rm in} + \omega)| = |\boldsymbol{d}_{\alpha}(\omega_{\rm in} - \omega)|$, giving no SA in $\bcal{E}_{\rm sc}$.
In general, BLS by magnetic excitations show classical SA~\cite{Wettling_SA, BLS_WGM_Mine}.

Using the collective coordinates approach discussed above, we find the $y$-component of the scattered field as,
\begin{equation}
    \boldsymbol{y} \cdot \bcal{E}_{\rm sc}(\boldsymbol{r}) \propto \text{Faraday} + \frac{if \mathcal{E}(\boldsymbol{0})  e^{-i\omega_{\rm in} t}}{|\boldsymbol{r}|} \left( \mathcal{S}_z - \mathcal{S}_{z0} \right). \label{Esc_y:cl}
\end{equation}
The first term describes a dominant elastic Faraday rotation, which carries negligible information about the skyrmion state. The second term assumes that the skyrmion is located at the origin, and is much smaller than the optical wavelengths. This is justified since optical wavelengths are in the micron range, while skyrmions are typically nanoscale. 

The $x$-component of the scattered field is,
\begin{equation}
    \boldsymbol{x}\cdot \bcal{E}_{\rm sc}(\boldsymbol{r}) \propto \frac{g\mathcal{E}(\boldsymbol{0})  e^{-i\omega_{\rm in} t}}{|\boldsymbol{r}|} \left[ \bar{S} \mathcal{S}_z -  \frac{\bar{S}^2 \mathcal{S}_z^2}{2 \mathcal{S}_{z0}^2} \sum_{\alpha} (1-\Pi_{0\alpha})^2 \right]. \label{Esc_x:cl}
\end{equation}

The above results show that the scattered light depends only on $\mathcal{S}_z$, and not on the helicity $\heli$.
If $\mathcal{S}_z$ were conserved, $\bcal{E}_{\rm sc}$ would have only one Fourier component at $\omega_{\rm in}$, i.e. no inelastic scattering.
Thus, BLS arises only when a symmetry-breaking potential of $\heli$ is present. As $\mathcal{S}_z$ is real, we get $\mathcal{S}_z(\omega) = \mathcal{S}_z^*(-\omega)$, implying sideband symmetry: $|E_{sc}(\omega_{\rm in} + \omega)| = |E_{sc}(\omega_{\rm in} - \omega)|$ for both $E_{sc} \in \{\boldsymbol{x}\cdot \bcal{E}_{\rm sc}, \boldsymbol{y}\cdot \bcal{E}_{\rm sc} \}$.
As we show below, this spectral symmetry is broken by vacuum fluctuations of the skyrmion.


~\\
\noindent \textbf{Quantum BLS by a skyrmion}. 
We now discuss the relation of the scattered quantized photons on quantum skyrmions. For simplicity we set $g=0$, neglecting linear birefringence, while the analysis is easily generalizable to $g\ne 0$. We present the salient features of the theory leaving details to the Appendix. The optical Hamiltonian is given by
\begin{equation} 
    \Ham{opt} = \hbar c \sum_{\sigma} \int_{-\infty}^{\infty} dk\ |k| \hat{a}_{\sigma,k}^{\dagger} \hat{a}_{\sigma,k} \,, 
\label{OptHam}
\end{equation}
where $\hat{a}_{\sigma,k}$ is the standard photon annihilation operator with polarization $\sigma \in \{x,y\}$ and wave-vector $k$.
The photon-skyrmion interaction leads to inter-polarization mixing via
\begin{equation}
	\hat{H}_{\rm int} = i \hbar \int_{-\infty}^{\infty} dkdk' g_{kk'} \left( \hat{a}_{x,k'}^{\dagger} \hat{a}_{y,k} - \hat{a}_{y,k'}^{\dagger} \hat{a}_{x,k} \right) \hat{\cal S}_z, \label{Def:Hint}
\end{equation}
with $g_{kk'} = \frac{f}{8\pi \hbar} \mathcal{E}_k \mathcal{E}_{k'}^*$ and $\mathcal{E}_{k}$ is the $k$-dependent electric field at the skyrmion center. 
To model spatial propagation, we define the forward and backward moving waves as
\begin{equation}
	\hat{a}_{\sigma \pm}(z,t) = \int_0^{\infty} \frac{dk}{\sqrt{2\pi}} e^{\pm ikz} \hat{a}_{\sigma,\pm k}(t). \label{def:forback}
\end{equation}
For an incoming $x$-polarized wave at frequency $\omega_{\rm in}$ and wave-vector $k_{\rm in} = \omega_{\rm in}/c$, we replace $\hat{a}_{x+}(z,t) \rightarrow a_{\rm in} e^{-i\omega_{\rm in}(t - z/c)}$, where $a_{\rm in}$ is the input amplitude.  

Since the skyrmion frequencies, Fig.~\ref{fig:energy}, lie in the $\si{\giga\hertz}$ range while optical frequencies are $>\SI{300}{\tera\hertz}$, scattering occurs only near $\omega_{\rm in}$, allowing the approximation $g_{kk_{\rm in}} \approx g_{k_{\rm in} k_{\rm in}}$ tantamount to assuming $\mathcal{E}_k \approx \mathcal{E}_{k_{\rm in}}$ for the relevant wave-vectors. 
Then, the Heisenberg equation yields the scattered photon field for $z>0$ and at times $t>z/c$ as $\hat{a}_{y+}(z,t) \equiv \hat{a}_{y+}(z-ct)$:
\begin{equation}
	\hat{a}_{y+}(z,t) = \hat{a}_{\rm vac}(z-ct) - \sqrt{\frac{\kappa_{\rm opt}}{c}} e^{-i\omega_{\rm in}(t - z/c)} \hat{\cal S}_z\left( t - \frac{z}{c} \right) . \label{ay:Markov} 
\end{equation}
The first term $\hat{a}_{\rm vac}(\Delta) \equiv \hat{a}_{y+}(\Delta,0)$ represents incoming vacuum noise in y-polarization. 
The second term shows that photons detected at time $t$ are radiated by the spins at retarded time $t-z/c$. The rate of photon generation from spins is proportional to
\begin{equation}
    \kappa_{\rm opt} = \left|\frac{f a_{\rm in} |\mathcal{E}_{k_{\rm in}}|^2}{4 \hbar \sqrt{c}} \right|^2, \label{Def:kopt}
\end{equation}
which has the dimensions of frequency. 

To detect the scattered photons, we consider an optical homodyne measurement with a local oscillator to select a frequency and amplify the signal.
It is sensitive to a typical photon operator at the detector location $z_{\rm det}$~\cite{Ulf},
\begin{equation}
	\hat{A}_{\omega} = \sqrt{c} \int dt\ \tilde{p}(t) e^{i(\omega_{\rm in} + \omega)t} \hat{a}_{y+}(z_{\rm det},t),
\end{equation}
with the normalization constraint $\int dt |\tilde{p}(t)|^2 = 1$. The photons emitted from the skyrmion at time $t$ will reach the detector at $t+z_{\rm det}/c$.
To detect the skyrmion radiation over a time-window of $(0,T)$ for some $T>0$, we put $\tilde{p}(t)$ to be $1/\sqrt{T}$ when $(t-z_{\rm det}/c) \in (0, T)$. 
We get the spectrum of the photon number as
\begin{equation}
    \langle \hat{A}_{\omega}^{\dagger} \hat{A}_{\omega} \rangle = \frac{\kappa_{\rm opt}}{T} \int_0^T dt dt' \langle \hat{\cal S}_z(t') \hat{\cal S}_z(t) \rangle e^{i\omega(t-t')}, \label{eq:OptCorr}
\end{equation}
proportional to the power spectral density of $\hat{\cal S}_z$. If $[\hat{\cal S}_z(t),\hat{\cal S}_z(t')] = 0$, then the BLS spectrum is sideband-symmetric $\langle \hat{A}_{\omega}^{\dagger} \hat{A}_{\omega} \rangle = \langle \hat{A}_{-\omega}^{\dagger} \hat{A}_{-\omega} \rangle$.
Thus, the quantum-mechanical nature of photons is not sufficient to produce SA if the skyrmion behaves classically.

~\\
\noindent \textbf{Skyrmion Dynamics.} To evaluate the BLS spectrum derived in Eq.~\eqref{eq:OptCorr}, we compute the time‐correlation of $\hat{\cal S}_z$ using open quantum-system methods~\cite{BP_OQS, RH_OQS}. Below, we outline the key steps and physical insights, with technical details provided in the Appendix. Dissipation of the skyrmion's quantum degrees of freedom occurs via two primary channels: photon emission, governed by $\hat{H}_{\rm int}$, and coupling to internal baths, such as magnons or impurities. When the bath forms a (quasi-)continuum and the dissipation is small, we can treat the skyrmion-bath interaction under a Markovian assumption, i.e., the information lost by the system to the bath is irretrievable~\cite{BP_OQS, RH_OQS}. 
Tracing out the bath produces the dynamics of the skyrmion's density matrix $\dot{\hat{\rho}} = \mathcal{L}\hat{\rho}$ in terms of a Liouvillian $\mathcal{L}$.

The dynamics of $\hat{\rho}$ can be separated into its diagonal and off-diagonal entries, written in terms of the skyrmion eigenmodes $\ket{u}$ and their energies $\hbar\omega_u$. The diagonal entries $p_u = \braket{u|\hat{\rho}|u}$ satisfy a master equation, $\dot{p}_u = \sum_v \left(\Gamma_{uv} p_v - \Gamma_{vu} p_u \right)$, where $\Gamma = \Gamma^{\rm opt} + \Gamma^{\rm int}$ includes both decay into the photons and the internal bath respectively. 
The off-diagonal components decay as 
$\braket{u|\hat{\rho}(t)|v} \sim e^{-i(\omega_u - \omega_v)t} e^{-\gamma_{uv} t}$, 
where $\gamma = \gamma^{\rm opt} + \gamma^{\rm int}$ is the rate of loss of quantum superposition between states. 
The steady-state density matrix defined by $\mathcal{L}\hat{\rho}_e = 0$ is diagonal, $\hat{\rho}_e = \sum_v p^e_u \ket{u} \bra{u}$, where $p_u^e$ are populations found from the master equation.

We now describe the dissipation matrices $\Gamma_{uv}$ and $\gamma_{uv}$. The optical dissipation terms are derived using $\Ham{int}$ defined in Eq.~(\ref{Def:Hint}), after tracing out the photons,
\begin{equation}
    \Gamma_{uv}^{\rm opt} = 2\kappa_{\rm opt} \left| \braket{u|\hat{\cal S}_z|v} \right|^2, \label{res:Gopt}
\end{equation}
where $\kappa_{\rm opt}$ given in Eq.~(\ref{Def:kopt}).
We note that $\Gamma^{\rm opt}$ tends to equalize the populations if $\braket{u|\hat{\cal S}_z|v} \ne 0$, effectively acting as an infinite-temperature bath. The optically-induced dephasing is
\begin{equation}
    \gamma_{uv}^{\rm opt} = \kappa_{\rm opt} \left( \braket{u|\hat{\cal S}_z^2|u} + \braket{v|\hat{\cal S}_z^2|v}  -2\braket{u|\hat{\cal S}_z|u} \braket{v|\hat{\cal S}_z|v} \right)\,. \label{res:gammaopt}
\end{equation}

The dissipation into internal degrees of freedom of the magnet depends on a given sample. We assume
\begin{equation}
    \Gamma_{uv}^{\rm int} =  \kappa_{\rm int}(\omega_v - \omega_u)\sum_k \left| \braket{u|\hat{D}_k|v}\right|^2\, ,
\end{equation}
where we choose the lowest-order dissipator terms $\hat{D}_k\in\{\hat{\cal S}_z, \cos\Opheli, \sin\Opheli\}$, and Ohmic damping consistent with the Gilbert phenomenology with rate $\kappa_{\rm int}(\omega) = \alpha_G |\omega|(n_{\rm BE} (|\omega|) + \Theta(\omega))$, with $\alpha_G$ the Gilbert damping, $n_{\rm BE}(\omega)$ the Bose-Einstein distribution, and $\Theta$ the Heaviside function. This form ensures detailed balance, $\kappa_{\rm int}(\omega) = \kappa_{\rm int}(-\omega) e^{\beta\omega}$ where $\beta = \hbar /k_BT$  and $T$ the temperature, implying relaxation toward thermal equilibrium,
\begin{equation}
    \lim_{t\rightarrow\infty} \left(\frac{p_v(t)}{p_u(t)}\right)_{\kappa_{\rm opt} = 0} = e^{-\beta (\omega_v - \omega_u)} \,.
\end{equation}
The dephasing due to internal dissipation sources is
\begin{equation}
    \gamma^{\rm int}_{uv} = \sum_e \frac{\Gamma^{\rm int}_{eu} + \Gamma^{\rm int}_{ev}}{2} - \kappa_{\rm int}(0) \sum_{k} \braket{u | \hat{D}_k | u} \braket{v| \hat{D}_k | v}, \label{res:gammaint}
    \end{equation}
where $\kappa_{\rm int}(0) = \alpha_G/\beta$. In the Appendix, we give the general form of the matrices $\Gamma^{\rm int}$ and $\gamma^{\rm int}$ for an arbitrary bath.

\begin{figure}[t]
    \centering
    \includegraphics[width=1\linewidth]{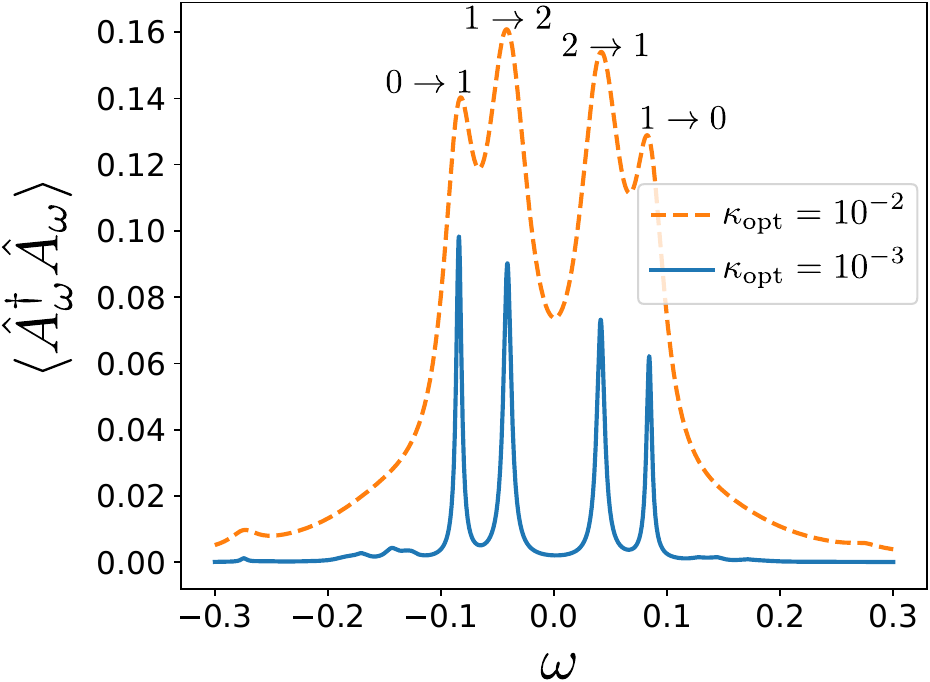}
    \caption{A typical BLS spectrum showing peaks at frequencies $\omega_{uv}$ corresponding to the $u\rightarrow v$ transition. For the plots, we choose the same values as Fig.~\ref{fig:energy} along with $B=1$, $\beta = 10$, and $\alpha_G = 10^{-2}$.}
    \label{fig:spec}
\end{figure}

We now compute the steady-state correlation function $C(\tau) = \lim_{t\rightarrow \infty} \langle \hat{\cal S}_z(t + \tau) \hat{\cal S}_z(t) \rangle$, which determines the emitted photon spectrum. Using the quantum regression formula~\cite{BP_OQS},
\begin{equation}
	C(\tau) = \mathrm{Tr}\left[
	\begin{cases}
		\hat{\cal S}_z e^{\mathcal{L}|\tau|} \left( \hat{\rho}_e \hat{\cal S}_z \right)  & \tau < 0 \\
		\hat{\cal S}_z e^{\mathcal{L}\tau} \left( \hat{\cal S}_z \hat{\rho}_e \right)  & \tau > 0
	\end{cases} \right] \,, \label{Corr:QRT}
\end{equation}
we obtain
\begin{equation}
    C(\tau) = \sum_{u\ne v} p_v \left| \braket{u|\hat{\cal S}_z|v} \right|^2 e^{-i(\omega_u -\omega_v)\tau} e^{-\gamma_{uv}|\tau|} + C_0(\tau) \,. \label{Res:CorrFun}
\end{equation}
This expression contains fast oscillating terms from coherences between eigenstates with $u\neq v$ and slowly varying terms $C_0(\tau)$ from population dynamics. Substituting into Eq.~\eqref{eq:OptCorr} we find the scattered photon number
\begin{equation}
    \langle \hat{A}_{\omega}^{\dagger} \hat{A}_{\omega} \rangle = \sum_{u \ne v} \frac{2\kappa_{\rm opt} p_v \left| \braket{u|\hat{\mathcal{S}}_z|v} \right|^2 \gamma_{uv}}{(\omega - \omega_{vu})^2 + \gamma^2_{uv}} + N_0(\omega), \label{res:OptSpec}
\end{equation}
where $\omega_{vu} = \omega_v - \omega_u$.
The low-frequency contribution $N_0(\omega)$ is on top of a dominant signal from Faraday rotation, and thus is hard to resolve experimentally. 

The BLS spectrum is a sum of Lorentzians with mean values at $\omega_{vu}$ and linewidth $\gamma_{uv}$, with $\gamma_{uv} = \gamma_{vu}$ by Hermiticity of density matrix. 
The peaks at $\omega_{vu}$ and $-\omega_{vu}$ have a ratio of $p_v/p_u$ and are generally different, giving an SA.
When optical dephasing is small, i.e. $\gamma^{\rm opt} \ll \gamma^{\rm int}$, this ratio is given by the Boltzmann factor $p_v/p_u = e^{-\beta(\omega_v - \omega_u)}$.
In contrast to classical BLS, a strong SA arises at low temperatures, which are experimentally accessible given that typical skyrmion energy gaps are a few GHz ($\SI{1}{\GHz} = \SI{50}{\milli\kelvin}\times k_B/h$).
The SA decreases with increasing optical power because photons induce heating, taking the ratio $p_v/p_u$ towards $1$. 

A typical BLS spectrum (without the zero frequency peak) is shown in Fig.~\ref{fig:spec}. 
For small values of external electric field, the transitions between adjacent levels are much stronger than other transitions.
This can be understood from the skyrmion Hamiltonian, Eq.~(\ref{Ham_skyr}), where the small perturbation $\propto \cos\Opheli$ hybridizes adjacent $\hat{\cal S}_z$ levels.
The peaks corresponding to $0\rightarrow1$ transitions are typically high due to a high population at $|0\rangle$, while those between $1\leftrightarrow 2$ transitions are high because of a higher matrix element $\braket{1|\hat{\cal S}_z|2}$. 
As optical power is increased, the peak amplitudes increase due to an increasing radiation by the skyrmion. 
However, SA decreases considerably as the population ratio $p_1/p_0$ and $p_2/p_1$ approach $1$.

\begin{figure}
    \centering
    \includegraphics[width=\linewidth]{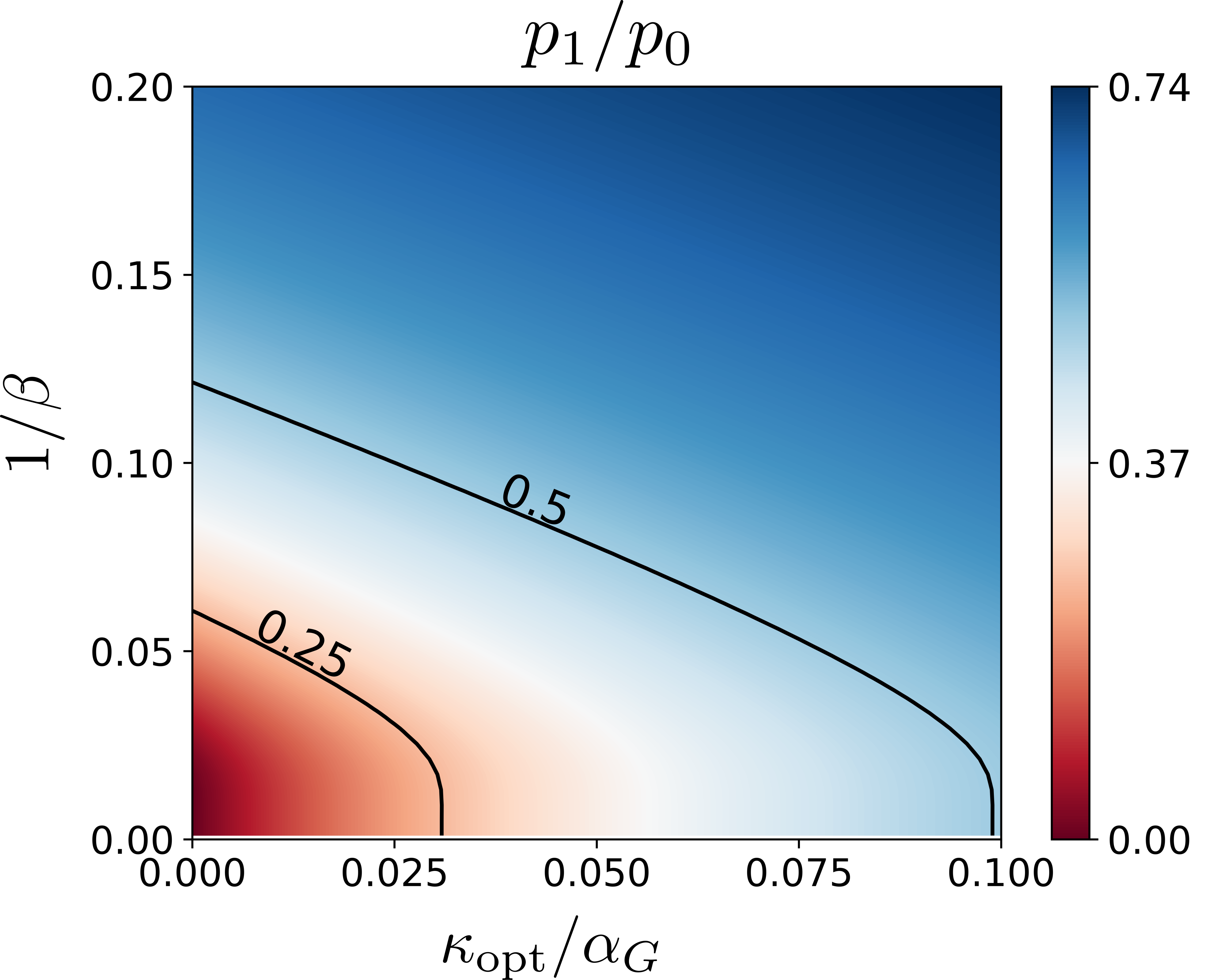}
    \caption{The ratio of $p_1/p_0$ as a function of optical dissipation and inverse temperature. 
    The parameters besides $\beta$ and $\kappa_{\rm opt}$ are the same as that used in Fig.~\ref{fig:spec}.} \label{fig:asymmetry}
\end{figure}

To quantify the asymmetry between the sidepeaks of $0\leftrightarrow 1$, we plot the ratio $p_1/p_0$ in Fig.~\ref{fig:asymmetry}, with respect to $\kappa_{\rm opt}$, proportional to the input laser power, and temperature $1/\beta$.
As either optical power or the temperature increases, $p_1/p_0$ gets closer to $1$, as expected.
However, there is a large window in the phase space where the ratio is $<0.5$ giving a large SA.

In conclusion, we showed that Brillouin light scattering, when applied to skyrmions in frustrated magnets, can reveal their quantum nature through sideband asymmetry. This is completely absent in the classical regime, but emerges from vacuum fluctuations and helicity level quantization.
A pronounced asymmetry ensures that quantum effects will dominate over small sources of classical asymmetries, such as skyrmion deformation.
While our analysis is applicable to other magnetic textures as well, particular care must be taken to choose a geometry where classical asymmetry vanishes.
Our analysis provides a novel practical protocol for optically probing quantum spin textures, and we expect to motivate experimental studies on their quantum control.

\newpage
\onecolumngrid
\appendix

\section{Collective coordinates of a skyrmion}

We consider a triangular lattice with competing interactions. Such lattices, for a given phase space, supports individual skyrmions over a ferromagnetic ground state. These skyrmions have an internal degree of freedom called the helicity. In this section, we review and extend the quantum theory of helicity and its conjugate momentum~\cite{SkyrQub_PRA}, the collective coordinates of the internal motion of a skyrmion. We review the path integral formalism for the spins in Sec.~\ref{subsec:PI_Quant}. Next, we describe the coordinate transformation from spins to collective coordinates in Sec.~\ref{subsec:coord_trans}. To get more insights into the coordinate transformation, we discuss their Poisson brackets in Sec.~\ref{subsec:Poisson}. Ignoring the magnons, we can use this coordinate transformation to find an effective Hamiltonian of only the collective coordinates, as discussed in Sec.~\ref{subsec:Sz_Ham}. Finally, we show the eigenvalue equation for the skyrmionic wave-function which forms a basis for all further analysis in Sec.~\ref{subsec:Eigens}.

\subsection{Path Integral Quantization} \label{subsec:PI_Quant}
We, first, review the path integral quantization of spins as a means to define the collective coordinates of the skyrmion. We consider a triangular lattice of spins $\{\vOp{S}_{\alpha}\}$, with maximum spin $\bar{S}$ and $\alpha$ being a lattice point. A typical spin-spin interaction is governed by the Hamiltonian,
\begin{equation}
    \hat{H} = \frac{1}{2} \sum_{\alpha \beta} J_{\alpha\beta} \vOp{S}_{\alpha} \cdot \vOp{S}_{\beta} - \frac{K}{2} \sum_{\alpha} \hat{S}_{\alpha z}^2 - B\sum_{\alpha} \hat{S}_{\alpha z},
\end{equation}
where the set $\{J_{\alpha\beta}\}$ are the exchange couplings, $K$ is the strength of an easy-axis anisotropy, and $B$ is proportional to an external magnetic field. 
Without loss of generality, we can assume $J_{\alpha\alpha} = 0$ as it does not have any effect on the dynamics and $J_{\alpha\beta} = J_{\beta\alpha}$ by redefining the coupling constants.
This Hamiltonian has a $U(1)$ symmetry of rotations around $z$-axis, i.e. with $\vOp{S}_{\alpha} \rightarrow e^{-i\Sz \heli} \vOp{S}_{\alpha} e^{i\Sz \heli}$, where the generator of rotation is
\begin{equation}
    \Sz = \sum_{\alpha} \left(\bar{S} - \hat{S}_{\alpha z} \right). \label{def:Sz_op}
\end{equation}
The subtraction from $\bar{S}$ is for consistency with definitions below.
Due to the azimuthal symmetry of the Hamiltonian, $\Sz$ is a conserved observable, i.e. $[\Sz, \hat{H}] = 0$.

The equivalent path integral formalism for this Hamiltonian is found using spin coherent states, 
\begin{equation}
    \ket{\{\Theta_{\alpha}, \Phi_{\alpha}\}} = \prod_{\alpha}  e^{-i\Phi_{\alpha} \hat{S}_{\alpha z}} e^{-i\Theta_{\alpha} \hat{S}_{\alpha y}} \ket{\text{FM}},
\end{equation}
where $\ket{\text{FM}}$ is the ferromagnetic ground state with all the spins pointing towards $+\boldsymbol{z}$.
Then, the amplitude between any two coherent states is given as a sum over paths with the corresponding Lagrangian,
\begin{equation}
    \mathcal{L} = \bar{S} \sum_{\alpha} (1 - \Pi_{\alpha})\dot{\Phi}_{\alpha} - H, 
\end{equation}
where $\Pi_{\alpha} = \cos\Theta_{\alpha}$ and the Hamiltonian is given by $H = \braket{\{\Theta_{\alpha}, \Phi_{\alpha}\}|\hat{H}|\{\Theta_{\alpha}, \Phi_{\alpha}\}}$.
With straightforward calculations, we find
\begin{equation}
    H = \frac{\bar{S}^2}{2} \sum_{\alpha\beta} J_{\alpha\beta} \boldsymbol{m}_{\alpha} \cdot \boldsymbol{m}_{\beta} - \frac{K \bar{S}(2\bar{S} - 1)}{4} \sum_{\alpha} \Pi^2_{\alpha} - B\bar{S} \sum_{\alpha} \Pi_{\alpha}, \label{def:PI_Ham}
\end{equation}
where
\begin{equation}
    \boldsymbol{m}_{\alpha} = \cos\Theta_{\alpha} \boldsymbol{z} + \sin\Theta_{\alpha} (\cos\Phi_{\alpha} \boldsymbol{x} + \sin\Phi_{\alpha} \boldsymbol{y}).
\end{equation}

The Berry phase term can also be written as $\bar{S} \sum_{\alpha} (1 - \cos\Theta_{\alpha})\dot{\Phi}_{\alpha} = \bar{S} \sum_{\alpha} \bcal{A}_{\alpha}\cdot \dot{\boldsymbol{n}}_{\alpha} $, where 
\begin{equation}
    \boldsymbol{n}_{\alpha} = \frac{\sqrt{1-\cos\Theta_{\alpha}}}{\sin\Theta_{\alpha}} \boldsymbol{m}_{\alpha},
\end{equation}
and $\bcal{A}_{\alpha} = \sqrt{1-\Pi_{\alpha}} \Apar_{\alpha} $, where
\begin{equation}
    \Apar_{\alpha} = -\sin\Phi_{\alpha} \boldsymbol{x} + \cos\Phi_{\alpha} \boldsymbol{y}.
\end{equation}

\subsection{Coordinate Transformation} \label{subsec:coord_trans}
To define the collective coordinates, we perform a coordinate transformation as described in this subsection. We, first, motivate the definitions of the new coordinates and later discuss the bijective transformation between the two coordinate systems.

Under certain parameter regimes, the `classical' Hamiltonian $H$ supports metastable skyrmions.
Consider a state $\{\Pi_{0\alpha}, \Phi_{0\alpha}\}$ or equivalently $\{\boldsymbol{n}_{0\alpha}\}$ which is a local minima of $H$. Due to the azimuthal symmetry mentioned above, $\mathcal{R}_{\heli}\boldsymbol{n}_{0\alpha}$ is also a minima where
\begin{equation}
    \mathcal{R}_{\heli} \boldsymbol{v} = v_z \boldsymbol{z} + \cos\heli \left(\boldsymbol{v} - v_z\boldsymbol{z}\right) + \sin\heli \boldsymbol{z}\times\boldsymbol{v},
\end{equation}
gives a rotation around $\boldsymbol{z}$ with the angle $\heli$. 
Thus, we can write $\boldsymbol{n}_{\alpha}$ in terms of a rotation and fluctuations on top of the skyrmion state,
\begin{equation}
    \boldsymbol{n}_{\alpha} = \mathcal{R}_{\heli} \left( \boldsymbol{n}_{0\alpha} + \boldsymbol{\gamma}_{\alpha} \right), \label{def:gamma}
\end{equation}
where $\boldsymbol{\gamma}_{\alpha}$ are fluctuations in the co-rotating frame, and $\heli$ is the helicity coordinate.

To completely specify the helicity dynamics, we also need to filter out the momentum of the helicity. We introduce fluctuations in the gauge vector around its ground state value $\bcal{A}_{0\alpha}$,
\begin{equation}
    \bcal{A}_{\alpha} = \mathcal{R}_{\heli} \left( \collcoor \bcal{A}_{0\alpha} + \boldsymbol{\Gamma}_{\alpha} \right), \label{def:Gamma}
\end{equation}
where $\collcoor$ is an abstract collective coordinate that will help set the momentum corresponding to the helicity, and $\boldsymbol{\Gamma}_{\alpha}$ are fluctuations in the gauge vector.
Clearly, $\boldsymbol{\gamma}_{\alpha}$ and $\boldsymbol{\Gamma}_{\alpha}$ are not independent vectors because $\bcal{A}_{\alpha} = \boldsymbol{z} \times \boldsymbol{n}_{\alpha}$. 
As the original theory involved only two spatially-dependent functions, $\{\Pi_{\alpha}, \Phi_{\alpha}\}$, the new coordinates should also involve only two functions. 
We define $\vs_{\alpha} = \boldsymbol{\gamma}_{\alpha} \cdot \Apar_{0\alpha}$, and $\vp_{\alpha} = \boldsymbol{\Gamma}_{\alpha} \cdot \Apar_{0\alpha}$, as these new coordinates, where $\Apar_{0\alpha}$ was defined above to be the direction of $\bcal{A}_{0\alpha}$.

As we have taken out two collective coordinates $\heli$ and $\collcoor$, we should also have two constraints to keep the number of degrees of freedom intact.
We choose the two constraints to be orthogonality of the fluctuations with the zero-energy state $\bcal{A}_{0\alpha}$, i.e.
\begin{equation}
    \sum_{\alpha} \bcal{A}_{0\alpha} \cdot \boldsymbol{\gamma}_{\alpha} = \sum_{\alpha} \bcal{A}_{0\alpha} \cdot \boldsymbol{\Gamma}_{\alpha} = 0. \label{constr:vec}
\end{equation}
These constraints can be equivalently written in terms of $\vs_{\alpha}$ and $\vp_{\alpha}$ as  $F_{\vs} = F_{\vp} = 0$, where
\begin{equation}
    F_X = \sum_{\alpha} \sqrt{1-\Pi_{0\alpha}} X_{\alpha},\ X\in\{\vs,\vp\}.
\end{equation}
These constraints are chosen to ensure that the Berry phase term separates out the fluctuations and collective coordinates,
\begin{equation}
    \bar{S} \sum_{\alpha} \bcal{A}_{\alpha} \cdot \dot{\boldsymbol{n}}_{\alpha} = \mathcal{S}_z \dot{\heli} + \bar{S} \sum_{\alpha} \boldsymbol{\Gamma}_{\alpha} \cdot \dot{\boldsymbol{\gamma}}_{\alpha},
\end{equation}
where we can also write $\sum_{\alpha} \boldsymbol{\Gamma}_{\alpha} \cdot \dot{\boldsymbol{\gamma}}_{\alpha} = \sum_{\alpha} \left( \vp_{\alpha} \dot{\vs}_{\alpha} - \vs_{\alpha} \dot{\vp}_{\alpha} \right)$, using $F_{\vs} = 0$.
Here, we defined
\begin{equation}
    \mathcal{S}_z = \bar{S} \sum_{\alpha} (1-\Pi_{\alpha}),
\end{equation}
as the classical counterpart of the conserved observable $\Sz$ defined in Eq.~(\ref{def:Sz_op}).

With this motivation, we consider the transformation
\begin{equation}
    \{ \Pi_{\alpha}, \Phi_{\alpha} \} \leftrightarrow \{\heli, \collcoor, \vs_{\alpha}, \vp_{\alpha}\}.
\end{equation}
By taking the dot product of Eqs.~(\ref{def:gamma}, \ref{def:Gamma}) with $\mathcal{R}_{\heli} \Apar_{0\alpha}$, we find
\begin{equation} \begin{aligned}
    \sqrt{1-\Pi_{\alpha}} \cos\left( \Phi_{\alpha} - \Phi_{0\alpha} - \heli \right) &= \collcoor \sqrt{1-\Pi_{0\alpha}} + \vp_{\alpha} \\
    \sqrt{1-\Pi_{\alpha}} \sin(\Phi_{\alpha} - \Phi_{0\alpha} - \heli) &= \vs_{\alpha}.
\end{aligned} \end{equation}
The above two equations can be succintly written as
\begin{equation}
    \sqrt{1-\Pi_{\alpha}} e^{i\Phi_{\alpha}} = \left( \vp_{\alpha} + i\vs_{\alpha} + \collcoor \sqrt{1-\Pi_{0\alpha}} \right) e^{i(\Phi_{0\alpha} + \varphi)},\label{res:coord_trans}
\end{equation}
which shows how to go from the new coordinates to the spin coordinates.

To find the reverse transformation, we use the constraints to find $\heli$ and $\collcoor$. Inserting the definitions of the fluctuations, $\boldsymbol{\gamma}_{\alpha}$ and $\boldsymbol{\Gamma}_{\alpha}$, into the constraints Eq.~(\ref{constr:vec}), we find
\begin{equation} \begin{aligned}
    0 &= \sum_{\alpha} \sqrt{1-\Pi_{0\alpha}} \sqrt{1-\Pi_{\alpha}} \sin(\Phi_{\alpha} - \Phi_{0\alpha} - \heli) \\
    0 &= \sum_{\alpha} \sqrt{1-\Pi_{0\alpha}} \sqrt{1-\Pi_{\alpha}} \cos(\Phi_{\alpha} - \Phi_{0\alpha} - \heli)  - \collcoor \frac{\mathcal{S}_{z0}}{\bar{S}}.
\end{aligned} \end{equation}
where 
\begin{equation}
    \mathcal{S}_{z0} = \bar{S} \sum_{\alpha} (1-\Pi_{0\alpha})
\end{equation}
is the ground state value of $\mathcal{S}_z$.
The above two constraints can be written in complex notation as
\begin{equation}
    \collcoor e^{i\heli}= \frac{\bar{S}}{S_{z0}} \sum_{\alpha} \sqrt{1-\Pi_{0\alpha}} \sqrt{1-\Pi_{\alpha}} e^{i(\Phi_{\alpha} - \Phi_{0\alpha})}, \label{collcoors_from_spins}
\end{equation}
defining the collective coordinates $\{\heli, \collcoor\}$, in terms of the spin coordinates.
Inserting these relations into Eq.~(\ref{res:coord_trans}), we can find $\vs_{\alpha}$ and $\vp_{\alpha}$ in terms of the spin coordinates.

Thus, we have completely described the bijective transformation between the two coordinate systems. 
While $\collcoor$ is useful to define as an intermediary quantity, it does not seem to have a simple physical interpretation. 
It is more useful to use the momentum of helicity $\mathcal{S}_z$ as a variable.
Thus, we change the coordinate system
\begin{equation}
    \{\heli, \collcoor, \vs_{\alpha}, \vp_{\alpha}\} \rightarrow \{\heli, \mathcal{S}_z, \vs_{\alpha}, \vp_{\alpha}\}.
\end{equation}
Taking absolute square on both sides of the coordinate transformation, Eq.~(\ref{res:coord_trans}) and using the constraint $F_{\vp}=0$, we find
\begin{equation}
    \mathcal{S}_z = \collcoor^2 \mathcal{S}_{z0} + \bar{S} \sum_{\alpha} (\vs_{\alpha}^2 + \vp_{\alpha}^2). \label{Sz_from_new}
\end{equation}
It appears that $\collcoor$ is the skyrmionic contribution to spin flips.

\subsection{Poisson Brackets} \label{subsec:Poisson}
To understand if the transformation is canonical, we calculate the Poisson brackets of the coordinates, defined by
\begin{equation}
    \{f,g\} = \frac{1}{\bar{S}} \sum_{\alpha} \left( \frac{\partial f}{\partial \Pi_{\alpha}} \frac{\partial g}{\partial \Phi_{\alpha}} - \frac{\partial f}{\partial \Phi_{\alpha}} \frac{\partial g}{\partial \Pi_{\alpha}} \right)
\end{equation}
First, we want to calculate the derivatives of the new coordinates in terms of the old ones.
For $\mathcal{S}_z$,
\begin{equation}
    \frac{\partial \mathcal{S}_z}{\partial \Pi_{\alpha}} = -\bar{S},\ \frac{\partial \mathcal{S}_z}{\partial \Phi_{\alpha}} = 0.
\end{equation}
For helicity, it is more convenient to write the derivatives of both $\collcoor$ and $\heli$ together using Eq.~(\ref{collcoors_from_spins}).
The two derivatives are,
\begin{equation} \begin{aligned}
    \frac{\partial \collcoor}{\partial \Pi_{\alpha}}  + i\collcoor \frac{\partial \heli}{\partial \Pi_{\alpha}} &= \frac{-\bar{S}}{2S_{z0}} \sqrt{\frac{1-\Pi_{0\alpha}}{1-\Pi_{\alpha}}} e^{i\xi_{\alpha}} \\
    \frac{\partial \collcoor}{\partial \Phi_{\alpha}}  + i\collcoor \frac{\partial \heli}{\partial \Phi_{\alpha}} &= \frac{i \bar{S}}{S_{z0}} \sqrt{1-\Pi_{0\alpha}} \sqrt{1-\Pi_{\alpha}} e^{i\xi_{\alpha}}, \label{Der_coll}
\end{aligned} \end{equation}
where $\xi_{\alpha} = \Phi_{\alpha} - \Phi_{0\alpha} - \heli $. 

Using the coordinate transformation, Eq.~(\ref{res:coord_trans}), we can find the derivatives of the rest of the coordinates,
\begin{equation} \begin{aligned}
    \frac{\partial (\vp_{\beta} + i\vs_{\beta})}{\partial \Pi_{\alpha}} &= \frac{-\delta_{\alpha\beta} e^{i\xi_{\beta}}}{2\sqrt{1-\Pi_{\beta}}} - i\sqrt{1-\Pi_{\beta}} e^{i\xi_{\beta}} \frac{\partial \heli}{\partial \Pi_{\alpha}} - \frac{\partial \collcoor}{\partial \Pi_{\alpha}} \sqrt{1-\Pi_{0\beta}} \\
    \frac{\partial (\vp_{\beta} + i\vs_{\beta})}{\partial \Phi_{\alpha}} &= i\sqrt{1-\Pi_{\beta}} e^{i\xi_{\beta}} \left(\delta_{\alpha\beta} - \frac{\partial \heli}{\partial \Phi_{\alpha}} \right) - \frac{\partial \collcoor}{\partial \Phi_{\alpha}} \sqrt{1-\Pi_{0\beta}}. \label{Der_newcoords}
\end{aligned} \end{equation}

From here on, the calculation of Poisson bracket is tedious, but straight-forward. The Poisson bracket of any function with $\mathcal{S}_z$ is
\begin{equation}
    \{ f, \mathcal{S}_z \} = \sum_{\alpha} \frac{\partial f}{\partial \Phi_{\alpha}}.
\end{equation}
Using Eq.~(\ref{Der_coll}), and the constraints $F_{\vp} + iF_{\vs} = 0$, we find
\begin{equation}
    \sum_{\alpha} \left( \frac{\partial \collcoor}{\partial \Phi_{\alpha}}  + i\collcoor \frac{\partial \heli}{\partial \Phi_{\alpha}} \right) = i\collcoor.
\end{equation}
Thus, $\{\collcoor, \mathcal{S}_z\}=0$ and $\{\heli, \mathcal{S}_z \} = 1$. 
The latter is consistent with $\mathcal{S}_z$ being conjugate to $\heli$.
Using Eq.~(\ref{Der_newcoords}), we can also find $\{\vp_{\beta} + i\vs_{\beta}, \mathcal{S}_z\} = 0$.

The rest of the Poisson brackets are derived after extensive algebra to be,
\begin{equation} \begin{aligned}
    \{\vp_{\beta} + i\vs_{\beta}, \heli\} &= 0 \\
    \{ \vp_{\alpha} + i\vs_{\alpha}, \vp_{\beta} + i\vs_{\beta} \} &= 0 \\
    \{\vp_{\alpha} + i\vs_{\alpha}, \vp_{\beta} - i\vs_{\beta} \} &= \frac{i}{\bar{S}} \left( \delta_{\alpha\beta} - \bar{S}\frac{\sqrt{1-\Pi_{0\alpha}} \sqrt{1-\Pi_{0\beta}}}{\mathcal{S}_{z0}} \right).
\end{aligned} \end{equation}

The first of the above relations shows independence of the helicity variable to the rest of the coordinates.
Without the term $\propto 1/\mathcal{S}_{z0}$, the above relations would imply that $\vs_{\alpha}$ and $\vp_{\alpha}$ form a set of canonical coordinates (with factors of $2\bar{S}$).
This last term is required because $\vs$ and $\vp$ are constrained.
For example, we know from constraints that $\sum_{\alpha} \sqrt{1-\Pi_{0\alpha}} (\vp_{\alpha} + i \vs_{\alpha}) = 0$, which turns out be consistent with the Poisson bracket,
\begin{equation}
    \sum_{\alpha} \sqrt{1-\Pi_{0\alpha}} \ \{\vp_{\alpha} + i\vs_{\alpha}, \vp_{\beta} - i\vs_{\beta} \} = 0.
\end{equation}
We did not attempt to form a set of canonical coordinates as it is not relevant for the rest of the manuscript.

\subsection{Hamiltonian} \label{subsec:Sz_Ham}
Using the coordinate transformation, Eq.~(\ref{res:coord_trans}), we can find the Hamiltonian in terms of the new coordinates. As we focus on the skyrmion dynamics, we put $\vs_{\alpha} = \vp_{\alpha} = 0$ to find $\Pi_{\alpha}$ and $\Phi_{\alpha}$. Inserting these values into the Hamiltonian, Eq.~(\ref{def:PI_Ham}), we find the Hamiltonian $H[\mathcal{S}_z]$. Note that due to the azimuthal symmetry, the Hamiltonian does not depend on the helicity $\heli$. 

We find this Hamiltonian by a numerical simulation of the triangular lattice with nearest and next-nearest neighbour exchange interactions with parameters $J_1$ and $J_2$. Within the parameters regime we considered, described in the main text, we can restrict the Hamiltonian to be,
\begin{equation}
    H[\mathcal{S}_z] \approx L(\mathcal{S}_z - \mathcal{S}_{z0})^2 + G(\mathcal{S}_z - \mathcal{S}_{z0})^3.
\end{equation}
Here, the Taylor coefficients $L$ and $G$ depend on the microscopic parameters $J_1$, $J_2$, $K$, and $B$.

\subsection{Eigenmodes with symmetry breaking} \label{subsec:Eigens}
So far, we have only considered a Hamiltonian with azimuthal symmetry giving the independence of $H$ on helicity, $\heli$.
If there are terms that break this symmetry, they will appear as potentials for helicity.
The usual correspondence between path integrals and quantum Hamiltonians suggest an effective Hamiltonian as,
\begin{equation}
\hat{H}_{\rm eff} = L(\hat{\cal S}_z - \mathcal{S}_{z0})^2 + G(\hat{\cal S}_z - \mathcal{S}_{z0})^3 + V(\Opheli),
\end{equation}
where $\Opheli$ and $\hat{\cal S}_z$ are the quantized counterparts of $\heli$ and $\mathcal{S}_z$, satisfying $[\hat{\heli},\hat{\cal S}_{z}]=i$ and $V$ is a potential for helicity.
From
\begin{equation}
\hat{\cal S}_{z}=\sum_{i}\left(\bar{S}-\hat{S}_{zi}\right),
\end{equation}
we find that $e^{2\pi i \hat{\cal S}_z} = 1$. 

On the space of periodic wave-functions $\{\psi | \psi(\heli + 2\pi) = \psi(\heli)\}$, we can also define a derivative operator $\hat{D} = -i\partial_{\heli}$, which satisfies the same relations as $\hat{\cal S}_z$: $[\hat{\heli},\hat{D}]=i$ and $e^{2\pi i\hat{D}} = 1$.
Thus, it allows us to choose $\hat{\cal S}_z - \hat{D}$ to be an integer.
For numerical accuracies in simulations, we choose $\hat{\cal S}_z - \lfloor S_{z0} \rfloor = \hat{D}$, where $\lfloor * \rfloor$ is the floor function, giving us an effective Hamiltonian 
\begin{equation}
    \hat{H}_{\rm eff} = L(i\partial_{\heli} + \delta\mathcal{S}_{z0})^2 - G(i\partial_{\heli} + \delta\mathcal{S}_{z0})^3 + V(\heli),
\end{equation}
where $\delta S_{z0} = S_{z0} - \lfloor \mathcal{S}_{z0} \rfloor$.
We find the eigenmodes of the above Hamiltonian by expanding the wave-functions in the exponential basis,
\begin{equation}
    \psi(\heli) = \sum_{n=-\infty}^{\infty} \alpha_n e^{in\heli},
\end{equation}
giving a matrix relation for $\alpha_n$.

\section{Optical Scattering by a skyrmion}

In this section, we review the derivation of the quantum Hamiltonian of the coupled system of traveling photons and a magnetic film holding a skyrmion. 
Using this Hamiltonian, we derive the frequency spectrum of the transmitted photons.
First, we discuss classical scattering of photons in Sec.~\ref{subsec:Class_Scat}. We, then, derive the classical Hamiltonian of the photon-skyrmion interaction in Sec.~\ref{subsec:OMag_Ham}. Finally, we discuss the quantum motion of the photons and find the output optical amplitude in Sec.~\ref{subsec:Opt_Out}.

\subsection{Classical Scattering} \label{subsec:Class_Scat}
To express the optical fields, electric $\boldsymbol{E}$, magnetic $\boldsymbol{B}$, and displacement $\boldsymbol{D}$, it is convenient to use the phasor notation such as $\boldsymbol{E} = \bcal{E} + \bcal{E}^*$ and similar for others. 
The classical electromagnetic Hamiltonian density is 
\begin{equation}
    \mathcal{H} = \frac{1}{2} \bcal{D}^*\cdot\bcal{E} + \frac{1}{2\mu_0} \bcal{B}^*\cdot\bcal{B}.
\end{equation}
Here, we assumed that the magnetic response at optical frequencies is negligible compared to the electric response, so the permeability is that of the vacuum.
Outside the magnet, the displacement field is given by $\bcal{D} = \varepsilon_0\bcal{E}$, giving a vacuum electromagnetic Hamiltonian density $\mathcal{H}_0$.

Inside the magnetic material, the displacement field depends on the magnetization.
For a continuum system with rotational symmetry, the general form of the displacement field is~\cite{FleuryLoudon},
\begin{equation}
	\bcal{D}(\boldsymbol{r}) = \varepsilon_r \varepsilon_0\bcal{E}(\boldsymbol{r}) + \sum_{\alpha} \delta(\boldsymbol{r}-\boldsymbol{r}_{\alpha}) \left[ if \bcal{E}(\boldsymbol{r})\times \boldsymbol{S}_{\alpha} + g \boldsymbol{S}_{\alpha} (\boldsymbol{S}_{\alpha}\cdot\bcal{E}(\boldsymbol{r})) \right].
\end{equation}
Here, $\varepsilon_r$ is the relative permittivity of the magnet, $f$ quantifies the Faraday rotation, and $g$ quantifies the Cotton-Mouton effect.

As a consequence, an incident field $\bcal{E}_{\rm in}$ induces a dipole moment given by $\boldsymbol{d} = \sum_{\alpha} \delta(\boldsymbol{r} - \boldsymbol{r}_{\alpha}) \boldsymbol{d}_{\alpha}$, where $\alpha$ are the lattice sites and
\begin{equation}
	\boldsymbol{d}_{\alpha} = if \boldsymbol{\cal E}_{\rm in}(\boldsymbol{r}_{\alpha})\times \boldsymbol{S}_{\alpha} + g \boldsymbol{S}_{\alpha} (\boldsymbol{S}_{\alpha}\cdot\boldsymbol{\cal E}_{\rm in}(\boldsymbol{r}_{\alpha})),
\end{equation}
with $f$ and $g$ quantifying circular and linear birefringence respectively.

At distances much greater than $k_{\rm in}r$, the far-field radiation moving in the $z$-direction is
\begin{equation}
    \bcal{E}_{\rm sc} \propto \sum_{\alpha} \frac{\boldsymbol{d}_{\alpha}^{\perp}}{|\boldsymbol{r} - \boldsymbol{r}_{\alpha}|} ,
\end{equation}
where $\boldsymbol{d}_{\alpha}^{\perp}$ is the component of the dipoles perpendicular to $z$.
For concreteness, we focus on a $x$-polarized monochromatic cylindrically symmetric input given by $\bcal{E}_{\rm in} = \mathcal{E}(\boldsymbol{\rho}) \boldsymbol{x} e^{i(kz - \omega_{\rm in}t)}$, where the transverse dependence of the amplitude $\mathcal{E}(\boldsymbol{\rho})$ varies at a length scale comparable to the optical wavelength $2\pi/k$. 
Then,
\begin{equation}
	\boldsymbol{d}_{\alpha}^{\perp} = \mathcal{E}(\boldsymbol{\rho}_{\alpha}) \left[ -if \boldsymbol{y} S_{\alpha, z} + g \boldsymbol{S}_{\alpha}^{\perp} S_{\alpha,x} \right] e^{i(kz_{\alpha} - \omega_{\rm in}t)}.
\end{equation}
If either $f$ or $g$ is zero, then clearly $\boldsymbol{d}_{\alpha}^{\perp}$ is sideband symmetric, i.e. $|\boldsymbol{d}_{\alpha}^{\perp}(\omega_{\rm in} + \omega)| = |\boldsymbol{d}_{\alpha}^{\perp}(\omega_{\rm in} - \omega)|$, implying that $\mathcal{E}_{\rm sc}$ is also sideband symmetric.
In general, however, this sideband symmetry will not hold.
We now look at the dipole moments due to skyrmion's helicity motion.

First, we look at $d_{\alpha y}$. The first term in the dipole moment, $\propto f$, gives a contribution (ignoring second order magnon terms),
\begin{equation}
    \sum_{\alpha} \frac{\mathcal{E}(\boldsymbol{\rho}_{\alpha})}{|\boldsymbol{r} - \boldsymbol{r}_{\alpha}|} \Pi_{\alpha} \approx \sum_{\alpha} \frac{\mathcal{E}(\boldsymbol{\rho}_{\alpha})}{|\boldsymbol{r} - \boldsymbol{r}_{\alpha}|} \left( \Pi_{0\alpha} - \frac{\mathcal{S}_z - \mathcal{S}_{z0}}{\mathcal{S}_{z0}} (1-\Pi_{0\alpha}) - 2 \vp_{\alpha} \collcoor \sqrt{1-\Pi_{0\alpha}} \right).
\end{equation}
The first term $\propto \Pi_{0\alpha}$ is a large elastic Faraday rotation. 
The second term $\propto \mathcal{S}_z$ gives a contribution from the internal degrees of freedom of the skyrmion and the last term, $\propto \vp_{\alpha}$, gives a contribution from magnons.
Because of the presence of the term $\sqrt{1-\Pi_{0\alpha}}$ only the magnon amplitude inside the skyrmion contributes to the magnon signal.
However, we expect that the magnons are spread over a large volume compared to the skyrmion.
Therefore, their amplitude inside the skyrmion will be much smaller than that of the skyrmion's internal states.
Thus, we ignore this contribution.
The fact that there is no linear contribution of the magnons in the ferromagnetic ground state in such a geometry is consistent with the selection rules applicable for a optomagnonic scattering in a uniform ground state magnetization.

Looking at the second term, $\propto \mathcal{S}_z - \mathcal{S}_{z0}$, we know from our simulations that $1-\Pi_{0\alpha}$ is non-zero only over a small distance of tens of atomic sites or lesser.
In such a small window, we can ignore the spatial dependence of $\mathcal{E}$ which spreads over a distance at least comparable to optical wavelengths.
Thus, we get
\begin{equation}
    \sum_{\alpha} \frac{\mathcal{E}(\boldsymbol{\rho}_{\alpha})}{|\boldsymbol{r} - \boldsymbol{r}_{\alpha}|} S_{\alpha z} \approx \text{Faraday} - \frac{\mathcal{E}(\boldsymbol{0}) e^{-i\omega_{\rm in} t}}{|\boldsymbol{r}|} \left( \mathcal{S}_z - \bar{S}\Lambda \right).
\end{equation}

The term $\propto g$, depends on $S_{\alpha y}S_{\alpha x}$, which in terms of the collective coordinates is,
\begin{equation}
    2S_{\alpha y} S_{\alpha x} = \bar{S}^2 (1+\Pi_{\alpha}) \mathrm{Im}\left[ \left(\vp_{\alpha} + i\vs_{\alpha} + \collcoor\sqrt{1-\Pi_{0\alpha}} \right)^2 e^{2i(\Phi_{0\alpha} + \heli)}\right].
\end{equation}
The terms linear in $\vs_{\alpha}$ and $\vp_{\alpha}$ appears multiplied by $\sqrt{1-\Pi_{0\alpha}}$, and thus, by a similar argument as above, they can be ignored.
For a circularly symmetry skyrmion, the summation over $\Phi_{0\alpha} = Q\phi_{\alpha}$ vanishes, giving no contribution.
Finally, we get
\begin{equation}
    \boldsymbol{y}\cdot\bcal{E}_{\rm sc} \propto \text{Faraday} + \frac{if \mathcal{E}(\boldsymbol{0})  e^{-i\omega_{\rm in} t}}{|\boldsymbol{r}|} \left( \mathcal{S}_z - \bar{S}\Lambda \right).
\end{equation}
As $\mathcal{S}_z$ is a real number, in frequency space, $\mathcal{S}_z(\omega) = \mathcal{S}_z^*(-\omega)$. 
This gives us a sideband symmetry, $|\boldsymbol{y}\cdot\bcal{E}_{\rm sc}(\omega_{\rm in} + \omega)| = |\boldsymbol{y}\cdot\bcal{E}_{\rm sc}(\omega_{\rm in} - \omega)|$.

Considering the $\boldsymbol{x}$-component of the dipole moment, $d_{\alpha x} \propto S_{\alpha x}^2$,
\begin{equation}
    2S_{\alpha x}^2 = \bar{S}^2 (1+\Pi_{\alpha} ) \mathrm{Re} \left[ (1-\Pi_{\alpha}) (1 + e^{2i\Phi_{\alpha}}) \right].
\end{equation}
By the same logic as before, the term containing $(1-\Pi_{\alpha}) e^{2i\Phi_{\alpha}} \propto e^{2\Phi_{0\alpha}}$ will vanish after a summation. The remaining term gives,
\begin{equation}
    \frac{2S_{\alpha x}^2}{\bar{S}^2} = 1-\left( 1 - \frac{\mathcal{S}_z}{\bar{S}} \frac{1-\Pi_{0\alpha}}{\Lambda} - 2\vp_{\alpha} \collcoor \sqrt{1-\Pi_{0\alpha}} \right)^2.
\end{equation}
Essentially the same reasoning as that in $d_{\alpha y}$ applies here as well, and we get
\begin{equation}
    \boldsymbol{x}\cdot \bcal{E}_{\rm sc} = \frac{g\mathcal{E}(\boldsymbol{0})  e^{-i\omega_{\rm in} t}}{|\boldsymbol{r}|} \left[ \bar{S} \mathcal{S}_z -  \frac{\bar{S}^2 \mathcal{S}_z^2}{2} \frac{\sum_{\alpha} (1-\Pi_{0\alpha})^2 }{\mathcal{S}_{z0}^2} \right].
\end{equation}
As everything inside the bracket is real, we again have sideband symmetry in $\boldsymbol{x}\cdot \bcal{E}_{\rm sc}$.

\subsection{Optomagnonic Hamiltonian} \label{subsec:OMag_Ham}
To obtain the quantum theory, we first find the classical optomagnonic Hamiltonian in terms of the collective coordinates of the skyrmion.
As we are interested only in the photons moving towards the $z$-axis, we can expand,
\begin{equation}
	\bcal{E}(\boldsymbol{\rho},z) = \sum_{\sigma} \int_{-\infty}^{\infty} \frac{dk}{\sqrt{2\pi}}\ \mathcal{E}_k(\boldsymbol{\rho}) e^{ikz} a_{\sigma,k} \boldsymbol{\sigma},
\end{equation}
Here, the position vector is decomposed as $\boldsymbol{r} = (\boldsymbol{\rho},z)$. 
$a_{\sigma,k}$ is the amplitude of the photons with the polarization $\sigma \in \{x,y\}$ and wave-vector $k$. 
$\mathcal{E}_k$ gives the dependence of the electric field in $\boldsymbol{\rho}$-direction.
The transverse spatial dependence of $\mathcal{E}_k$ depends on the particular setup of optical focussing and the optical excitation, and will typically be comparable to the optical wavelength.
A typical example would be a Gaussian profile with width at least comparable to the wavelength.

A similar expansion applies to the magnetic field of the photons. 
This simplifies the optical Hamiltonian without the magneto-optical effect,
\begin{equation}
	H_{\rm opt} = \frac{1}{4} \int dV \left[ \varepsilon_0 |\bcal{E}|^2 + \frac{|\bcal{B}|^2}{\mu_0} \right] = \sum_{\sigma} \int_{-\infty}^{\infty} dk\ E_k |a_{\sigma,k}|^2,
\end{equation}
 where we used standard orthogonality relations between optical modes, and the energy of each mode is,
\begin{equation}
	E_k = \frac{\varepsilon_0}{2} \int |\mathcal{E}_k(\boldsymbol{\rho})|^2 dx dy.
\end{equation}
Here, we used equipartition theorem to eliminate the term containing the magnetic field.
This leads to an interaction Hamiltonian, $H_{\rm int} = H_f + H_g$, where
\begin{equation}
    H_f = \frac{if}{4} \sum_{\alpha} \int \frac{dk dk'}{2\pi} \mathcal{E}_k(\boldsymbol{\rho}_{\alpha}) \mathcal{E}_{k'}^*(\boldsymbol{\rho}_{\alpha}) e^{i(k-k')z_{\alpha}} \left( a_{x,k} a^*_{y,k'} - a_{y,k} a^*_{x,k'} \right) S_{\alpha z}.
\end{equation}
and 
\begin{equation}
    H_g = \frac{g}{4} \sum_{\alpha, \sigma, \sigma'} \int \frac{dk dk'}{2\pi}\ \mathcal{E}_{k'}^*(\boldsymbol{\rho}_{\alpha}) \mathcal{E}_k(\boldsymbol{\rho}_{\alpha}) e^{i(k-k')z_{\alpha}} a^*_{\sigma',k'}  a_{\sigma,k} S_{\alpha, \sigma'} S_{\alpha, \sigma}
\end{equation}
In the previous section, we saw that besides an elastic Faraday rotation, we can treat the magneto-optical effect by assuming the skyrmion as a point-dipole.
With the same approximation, we find,
\begin{equation}
    H_f = \text{Faraday} - \frac{if}{4} \int \frac{dk dk'}{2\pi} \mathcal{E}_k(\boldsymbol{0}) \mathcal{E}_{k'}^*(\boldsymbol{0}) \left( a_{x,k} a^*_{y,k'} - a_{y,k} a^*_{x,k'} \right) \mathcal{S}_z.
\end{equation}
and
\begin{equation}
    H_g = \frac{g}{4} \sum_{\alpha, \sigma} \int \frac{dk dk'}{2\pi}\ \mathcal{E}_{k'}^*(\boldsymbol{0}) \mathcal{E}_k(\boldsymbol{0}) a^*_{\sigma,k'}  a_{\sigma,k} \left[ \bar{S} \mathcal{S}_z -  \frac{\bar{S}^2\mathcal{S}_z^2}{2} \frac{\sum_{\alpha} (1-\Pi_{0\alpha})^2 }{S_{z0}^2} \right].
\end{equation}
For brevity in expressions, we analyze the case with $g=0$ below.

\subsection{Quantum dynamics} \label{subsec:Opt_Out}

To quantize the classical Hamiltonian, we promote the photon amplitudes $a_{\sigma,k} \rightarrow \hat{a}_{\sigma, k}$ where $\hat{a}_{\sigma,k}$ is the annihilation operator of the photons with the polarization $\sigma \in \{x,y\}$ and wave-vector $k$ satisfying the field commutation relations, $[\hat{a}_{\sigma,k},\hat{a}_{\sigma',k'}^{\dagger}] = \delta_{\sigma\sigma'}\delta(k-k')$. 
Choosing the energy of each mode to be that of a photon, $E_k = \hbar c |k|$, we get the quantum optical Hamiltonian,
\begin{equation}
	\hat{H}_{\rm opt} = \hbar \sum_{\sigma} \int_{-\infty}^{\infty} dk\ c|k| \hat{a}_{\sigma,k}^{\dagger} \hat{a}_{\sigma,k}.
\end{equation}
As discussed in the classical dynamics, we can treat the skyrmion as a point-dipole. 
Quantizing the above classical Hamiltonian,  
\begin{equation}
	\hat{H}_{\rm int} \approx i \hbar \int_{-\infty}^{\infty} dkdk' g_{kk'} \left( \hat{a}_{x,k'}^{\dagger} \hat{a}_{y,k} - \hat{a}_{y,k'}^{\dagger} \hat{a}_{x,k} \right) \Sz,
\end{equation}
where the coupling constant is
\begin{equation}
    g_{kk'} = \frac{f \mathcal{E}_k(\boldsymbol{0}) \mathcal{E}_{k'}^*(\boldsymbol{0}) }{8\pi \hbar}.
\end{equation}

Similar to the classical dynamics, we assume an $x$-polarized input, and a $y$-polarized output. We solve for the output photons via their equation of motion,
\begin{equation}
	\frac{d \hat{a}_{y,k}}{dt} = -ic|k|\hat{a}_{y,k} - \int dk' g_{k'k} \hat{a}_{x,k'} \Sz.
\end{equation}
Consider a time $t_0$, long before the photons interact with the spins. Solving the above equation starting from $t_0$,
\begin{equation}
	 \hat{a}_{y,k}(t) = \hat{a}_{y,k}(t_0)e^{-i|k|c(t-t_0)} - \int_{t_0}^t d\tau \int dk'g_{kk'} \hat{a}_{x,k'}(\tau) \Sz(\tau) e^{-i|k|c(t-\tau)}. \label{Sol:ayk}
\end{equation}

We convert the above solution to real-space separating forward and backward moving waves, $\hat{a}_y(z,t) = \hat{a}_{y+}(z,t) + \hat{a}_{y-}(z,t)$, where each is an integral over only the positive or negative wave-vectors,
\begin{equation}
	\hat{a}_{y\pm}(z,t) = \int_0^{\infty} \frac{dk}{\sqrt{2\pi}} e^{\pm ikz} \hat{a}_{y,\pm k}(t).
\end{equation}

For brevity, we consider only the transmitted field as the reflected field carries nearly the same information. To derive transmission of photons, we will focus on the photonic component moving forwards, given by
\begin{equation}
	\hat{a}_{y+}(z,t) = \hat{a}_{y+}(z-c(t-t_0), t_0) - \int_0^{\infty} \frac{dk}{\sqrt{2\pi}} \int_{t_0}^t d\tau \int dk'g_{kk'} \hat{a}_{x,k'}(\tau) \Sz(\tau) e^{ik(z-c(t-\tau))}.
\end{equation}
The first term describes a free evolution of photons moving at speed $c$, while the second term describes the optomagnonic scattering by the skyrmion.
Converting $\hat{a}_{x,k'}$ to position,
\begin{equation}
	\hat{a}_{y+}(z,t) = \hat{a}_{y+}(z-c(t-t_0), t_0) - \int_{t_0}^t d\tau \int \frac{dz'}{\sqrt{2\pi}} \left[\int_0^{\infty} \frac{dk}{\sqrt{2\pi}}  \int dk'g_{kk'} e^{-ik'z'} e^{ik(z-c(t-\tau))}\right] \hat{a}_{x}(z',\tau) \Sz(\tau) .
\end{equation}

This is valid for a general input in $x$-polarization. Now, we assume a classical input at a frequency $\omega_{\rm in}$ and a wave-vector $k_{\rm in} = \omega_{\rm in}/c$, which allows us to replace $\hat{a}_x(z,t) \rightarrow a_{\rm in} e^{-i\omega_{\rm in}(t - z/c)}$ where $a_{\rm in}$ is the amplitude of the input wave. 
The transmitted photon amplitude becomes
\begin{equation}
	\hat{a}_{y+}(z,t) = \hat{a}_{y+}(z-c(t-t_0), t_0) - a_{\rm in} \int_{t_0}^t d\tau \left[\int_0^{\infty} dk g_{kk_{\rm in}} e^{ik(z-c(t-\tau))}\right] e^{-i\omega_{\rm in}\tau} \Sz(\tau) .
\end{equation}
Optical frequencies are 5-6 orders of magnitude higher than magnonic frequencies, so by energy conservation, the scattering is significant only when $k\approx k_{\rm in}$. So, we approximate $g_{kk_{\rm in}} \approx g_{k_{\rm in} k_{\rm in}}$ independent of $k$. Thus,
\begin{equation}
	\hat{a}_{y+}(z,t) = \hat{a}_{y+}(z-c(t-t_0), t_0) - \frac{f a_{\rm in} |\mathcal{E}_{k_{\rm in}}(\boldsymbol{0})|^2}{4 \hbar c} e^{-i\omega_{\rm in}(t - z/c)} \Sz\left( t - \frac{z}{c} \right) .
\end{equation}
In removing the integral w.r.t $\tau$, we have tacitly assumed $t-z/c > t_0$ which holds for $t_0$ being far back in the past. As expected from wave motion, the spin at time $t$ affects the photons at $z>0$ at time $t+z/c$.

The wave at $t_0 \rightarrow -\infty$ before scattering by skyrmion is just vacuum,
\begin{equation}
	\hat{a}_y(z-c(t-t_0), t_0) = \hat{a}_{\rm vac}(z-ct),
\end{equation}
where the vacuum operator satisfies
\begin{equation}
	\langle \hat{a}_{\rm vac}^{\dagger}(\xi) \hat{a}_{\rm vac}(\xi') \rangle = 0.
\end{equation}
Note the field commutation relation $[\hat{a}_{\rm vac}(\xi), \hat{a}_{\rm vac}^{\dagger}(\xi')] = \delta(\xi - \xi')$, that can be derived from the commutation relations of $\hat{a}_{yk}$ written above.

A typical optical measurement involves a filter to select a frequency followed by a photo-detector, giving an output photon
\begin{equation}
	\hat{A}_{\omega} = \sqrt{c} \int dt\ \tilde{p}(t) e^{i(\omega_{\rm in} + \omega)t} \hat{a}_{y+}(z_{\rm det},t),
\end{equation}
where $\tilde{p}(t)$ is a filter function which is zero outside the measurement window, $z_{\rm det}$ is the detector location, and we are measuring the frequency relative to the input frequency $\omega_{\rm in}$. The measured operator is that of a photon number, and thus satisfy $[\hat{A}_{\omega}, \hat{A}_{\omega}^{\dagger}] = 1$ equivalent to
\begin{equation}
	\int dt |\tilde{p}(t)|^2 = 1
\end{equation}
The output can be divided into a noise and a signal component, $\hat{A}_{\omega} = \hat{N}_{\omega} - \hat{S}_{\omega}$, where
\begin{equation}\begin{aligned}
	\hat{N}_{\omega} &= \sqrt{c} \int dt\ p(t) e^{i(\omega_{\rm in} + \omega)t} \hat{a}_{\rm vac}(- ct), \\
	\hat{S}_{\omega} &= \frac{f a_{\rm in} |\mathcal{E}_{k_{\rm in}}|^2}{4 \hbar \sqrt{c}} \int dt\ p(t) e^{i\omega t} \Sz(t)
\end{aligned}\end{equation}
We defined $p(t) = \tilde{p}(t+z_{\rm det}/c) e^{i(\omega_{\rm in} + \omega) z_{\rm det}/c}$ which satisfies the same normalization as $\tilde{p}(t)$.

The number of photons for each $\omega$ is given by $\langle \hat{A}_{\omega}^{\dagger} \hat{A}_{\omega} \rangle$. To calculate such averages, we assume an initial density matrix that is separable, $\hat{\rho} = \hat{\rho}_{\rm mag} \otimes \hat{\rho}_{\rm opt}$. In the Heisenberg picture, $\hat{\rho}$ does not change with time. So, at all times, $\hat{\rho}_{\rm opt}$ contains no photons in $y$-polarization, giving $\hat{N}_{\omega} \hat{\rho} = \hat{\rho} \hat{N}_{\omega}^{\dagger} = 0$. This gives a simple result 
\begin{equation}
    \langle \hat{A}_{\omega}^{\dagger} \hat{A}_{\omega} \rangle = \langle \hat{S}_{\omega}^{\dagger} \hat{S}_{\omega} \rangle .
\end{equation}
Note that the noise cannot be eliminated at higher orders.

Consider a case of a fixed measurement window such that $p(t) = 1/\sqrt{T}$ within $(0, T)$ and zero otherwise. Writing it out in time domain,
\begin{equation}
    \langle \hat{S}_{\omega}^{\dagger} \hat{S}_{\omega} \rangle = \frac{\kappa_{\rm opt}}{T} \int_0^T dt dt' \langle \Sz(t') \Sz(t) \rangle e^{i\omega(t-t')},
\end{equation}
where we defined
\begin{equation}
    \kappa_{\rm opt} = \left|\frac{f a_{\rm in} |\mathcal{E}_{k_{\rm in}}|^2}{4 \hbar \sqrt{c}} \right|^2, \label{Def:kappaopt}
\end{equation}
which can be interpreted as the rate of photon generation by spins.

The difference between $\omega$ and $-\omega$ is
\begin{equation}
	\langle \hat{S}_{\omega}^{\dagger} \hat{S}_{\omega} \rangle - \langle \hat{S}_{-\omega}^{\dagger} \hat{S}_{-\omega} \rangle = \frac{\kappa_{\rm opt}}{T} \int_0^T dt dt' \left\langle \left[\Sz(t'), \Sz(t) \right] \right\rangle e^{i\omega(t-t')}.
\end{equation}
Classically, $S_z(t) S_z(t') = S_z(t') S_z(t)$ and thus, the difference vanishes. Quantum mechanically, this is not true in general.

\section{Skyrmion Dynamics}
Here, we discuss the dissipative dynamics of the skyrmion with the purpose of calculating the power spectrum of $\hat{\mathcal{S}}_z$ which ultimately gives the output optical spectrum. 
In Sec.~\ref{ssec:Lind}, we review the general theory of Lindblad equation from a generic microscopic system-bath Hamiltonian. 
In Sec.~\ref{ssec:Opt} and \ref{ssec:Int}, we specialize this general theory to the case of dissipation due to optical photons and internal bath respectively. 
In Sec. \ref{ssec:Spec}, we use the ensuing equations of motion to derive the power spectrum of $\Sz$.

\subsection{General Lindblad Formalism} \label{ssec:Lind}
Here, we review the results from the general theory of open quantum systems to derive an effective Markovian equation of motion for a dissipative system.
Consider a system with the Hamiltonian $\hat{H}_S = \hbar\sum_{n} \omega_n \ket{n}\bra{n}$ decomposed into eigenmodes $\ket{n}$ and corresponding eigenfrequencies $\omega_n$ with a non-negative integer index $n\in\mathbb{Z}_+$. 
The number of modes can be finite or infinite.
Dissipation can be modeled by a coupling of the system to a continuum of states forming a bath. 
Let the bath Hamiltonian be  $\hat{H}_B = \hbar \int d\alpha\, \omega_{\alpha} \ket{\alpha}\bra{\alpha}$ again decomposed into eigenmodes $\ket{\alpha}$ and eigenfrequencies $\omega_{\alpha}$. 
A general system-bath coupling can be written as $\hat{H}_I = \sum_k \hat{D}_k \otimes \hat{B}_k$, where $\hat{D}_k$ is a system operator, $\hat{B}_k$ is a bath operator, and the index $k$ can be countable or a continuum.
By separation into Hermitian and non-Hermitian components, if necessary, we can always assume $\hat{D}_k$ and $\hat{B}_k$ to be Hermitian. 
In that case, the reduced dynamics of the system under Markovian assumption is given by~\cite{BP_OQS, RH_OQS}
\begin{equation}
    \frac{d \hat{\rho}}{dt} = -i[\hat{H}_S, \hat{\rho}] + \sum_{\eta, kl} \kappa_{kl}(\eta) \left[ \hat{D}_l(\eta) \hat{\rho} \hat{D}_k(-\eta) - \frac{1}{2} \left\{ \hat{D}_k(-\eta) \hat{D}_l(\eta), \hat{\rho} \right\} \right] \label{eq:appLind}
\end{equation}
Here, $\hat{\rho}$ is the density matrix of the system. 
We have absorbed a small Lamb shift of the energy levels into the system Hamiltonian $\hat{H}_S$.
The variable $\eta$ spans the frequency differences between system levels, i.e., over the set $\eta \in \{\omega_m - \omega_n | m,n\in \mathbb{Z}_+ \}$. 
The Lindblad operators above are defined via the spectral decomposition, 
\begin{equation}
	\hat{D}_k(\eta) = \sum_{\omega_m - \omega_n = \eta} \ket{n}\braket{n|\hat{D}_k|m} \bra{m}
\end{equation}
involving a sum over all pairs of levels with frequency difference $\eta$. 
The dissipative coefficients are given by
\begin{equation}
    \kappa_{kl}(\eta) = 2\pi \mathrm{Tr}\left[\hat{B}_k(\eta)\hat{B}_l\hat{\rho}_B \right], \label{def:kappakl}
\end{equation}
where $\hat{B}_k(\eta)$ are again given by a spectral decomposition but on the bath modes, 
\begin{equation}
    \hat{B}_k(\eta) = \int d\alpha d\beta\ \delta(\omega_{\beta} - \omega_{\alpha} - \eta) \ket{\alpha} \braket{\alpha|\hat{B}_k|\beta} \bra{\beta}
\end{equation}
and $\hat{\rho}_B$ is the bath's density matrix. 

A quick method to identify the spectral decomposition of an operator is to look at their time evolution without interaction, e.g. for a system operator,
\begin{equation}
    e^{i\hat{H}_S t} \hat{D}_k(\eta) e^{-i\hat{H}_S t} = e^{-i\eta t} \hat{D}_k(\eta).
\end{equation}
As $\hat{D}_k = \sum_\eta \hat{D}_k(\eta)$, we have
\begin{equation}
    e^{i\hat{H}_S t} \hat{D}_k e^{-i\hat{H}_S t} = \sum_{\eta} e^{-i\eta t} \hat{D}_k(\eta).
\end{equation}
This implies that one can read-off $\hat{D}_k(\eta)$ from the time evolution (without interaction) of $\hat{D}_k$ as the coefficient of $e^{-i\eta t}$.
A similar result holds for the bath operators,
\begin{equation}
    e^{i\hat{H}_B t} \hat{B}_k e^{-i\hat{H}_B t} = \int d\eta e^{-i\eta t} \hat{B}_k(\eta).
\end{equation}
We use this property below to find the spectral decomposition of optically-induced bath operators.

We unravel the above Lindblad equation, Eq.~(\ref{eq:appLind}), for a generic density matrix term, $\rho_{uv} = \braket{u | \hat{\rho} | v}$, where $\ket{u}$ and $\ket{v}$ are system eigenstates.
Its equation of motion is
\begin{equation}
    \frac{d \rho_{uv}}{dt} = -i(\omega_u - \omega_v)\rho_{uv} + \sum_{\eta, kl} \kappa_{kl}(\eta) \Braket{u | \left[ \hat{D}_l(\eta) \hat{\rho} \hat{D}_k(-\eta) - \frac{1}{2} \left\{ \hat{D}_k(-\eta) \hat{D}_l(\eta), \hat{\rho} \right\} \right] | v } \label{eq:rho_uv}
\end{equation}
Consider the first dissipative term,
\begin{equation}
	\braket{u|\hat{D}_l(\eta) \hat{\rho} \hat{D}_k(-\eta)|v} = \sum_{\omega_{\alpha} - \omega_u = \eta} \sum_{\omega_v - \omega_{\beta} = -\eta} \braket{u|\hat{D}_l|\alpha} \braket{\beta|\hat{D}_k|v} \rho_{\alpha\beta} \label{eq:LindTerm}
\end{equation}
In the above summation, for any $\eta$, we have $\omega_{\alpha} - \omega_{\beta} = \omega_u - \omega_v $. 
The RHS depends only on those $\{\alpha,\beta\}$ that satisfy this difference condition. 
Similarly, the second dissipative term is,
\begin{equation}
    \Braket{u | \hat{D}_k(-\eta) \hat{D}_l(\eta) \hat{\rho} | v} = \sum_{\omega_{\alpha} - \omega_{\beta} = \eta} \sum_{\omega_{\beta} - \omega_u = -\eta} \braket{u | \hat{D}_k | \beta} \braket{\beta| \hat{D}_l |\alpha} \hat{\rho}_{\alpha v} 
\end{equation}
Here, we find a restriction on the summation as $\omega_{\alpha} = \omega_u$.
A similar condition can be derived for the final dissipative term in Eq.~(\ref{eq:rho_uv}). 
Using all of the above, we can write,
\begin{equation}
    \frac{d \rho_{uv}}{dt} = -i(\omega_u - \omega_v)\rho_{uv} - \sum_{\omega_{\alpha} - \omega_{\beta} = \omega_u - \omega_v} \gamma_{uv,\alpha\beta} \rho_{\alpha\beta} \label{rhouv:div}
\end{equation}
We can divide the density matrix into sectors of a given frequency difference $\eta$, $\{\rho_{uv} | \omega_u - \omega_v = \eta\}$. 
Each of these sectors can be diagonalized independent of each other.
The $\eta = 0$ sector includes the diagonal entries.

While it is possible to proceed further with the general case, the notation becomes very cumbersome. 
For simplicity, we restrict ourselves to the two assumptions as mentioned in the main text. 
First is the lack of degeneracy, $\omega_n \neq \omega_m$ for $n \ne m$. 
Second is to have no spectral degeneracy, $\omega_n - \omega_m \ne \omega_p - \omega_q$ unless $n=p$ and $m=q$. 

For diagonal entries, $p_u\equiv \rho_{uu}$, we need to consider only the density matrix elements ${\rho}_{\alpha\beta}$  with $\omega_{\alpha} = \omega_{\beta}$ in Eq.~(\ref{rhouv:div}). 
With no degeneracy, this implies $\alpha = \beta$. 
Thus, we can write an equation of motion for only the diagonal entries using Eq.~(\ref{eq:rho_uv}) as a detailed balance master equation,
\begin{equation}
    \dot{p}_u = \sum_v \Gamma_{uv} p_v - \left(\sum_v \Gamma_{vu}\right) p_u \label{evol:prob_u}
\end{equation}
where
\begin{equation}
	\Gamma_{uv} = \sum_{kl} \kappa_{kl}(\omega_v - \omega_u) \braket{u | \hat{D}_l | v} \braket{v | \hat{D}_k | u}. \label{res:gen_Gamma}
\end{equation}
The above master equation satisfies the condition of probability conservation $\sum_u \dot{p}_u = 0$.

Under spectral non-degeneracy, for off-diagonal entries $\rho_{uv}$ with $u\ne v$, the summation in Eq.~(\ref{rhouv:div}) reduces to only $\alpha = u$ and $\beta = v$. Thus, we can write
\begin{equation}
    \frac{d \rho_{uv}}{dt} = -i(\omega_u - \omega_v)\rho_{uv} - \gamma_{uv} \rho_{uv} \label{evol:off_diag}
\end{equation}
defining $\gamma_{uv}$ as the effective dephasing rate for this off-diagonal entry,
\begin{equation}
    \gamma_{uv} = -\sum_{kl} \kappa_{kl}(0) \braket{u | \hat{D}_l | u} \braket{v| \hat{D}_k | v} + \sum_e \frac{\Gamma_{eu} + \Gamma_{ev}}{2}. \label{res:gen_gamma}
\end{equation}
In the following, we calculate $\gamma_{uv}$ and $\Gamma_{uv}$ for the optical and internal dissipation.

\subsection{Optical Dissipation}\label{ssec:Opt}
In case of optical dissipation, the interaction is given by Eq.~(\ref{Def:Hint}), where we have only one skyrmion operator $\hat{D}_0 = \Sz$ with the optical operator
\begin{equation}
    \hat{B}_0 = i \int dkdk' g_{kk'} \left( \hat{a}_{x,k'}^{\dagger} \hat{a}_{y,k} - \hat{a}_{y,k'}^{\dagger} \hat{a}_{x,k} \right).
\end{equation}
To calculate the spectral decomposition, consider the free evolution of $\hat{B}_0$,
\begin{equation}
    e^{i\hat{H}_Bt} \hat{B}_0 e^{-i\hat{H}_Bt} = i \int dkdk' e^{ict(|k'| - |k|)} g_{kk'} \left( \hat{a}_{x,k'}^{\dagger} \hat{a}_{y,k} - \hat{a}_{y,k'}^{\dagger} \hat{a}_{x,k} \right),
\end{equation}
where we used $e^{i\hat{H}_B t} \hat{a}_{\sigma, k} e^{-i\hat{H}_B t} = e^{-ic |k| t}\hat{a}_{\sigma,k}$.
As discussed above, the spectral decomposition is the component of $e^{-i\eta t}$, i.e.,
\begin{equation}
    \hat{B}_0(\eta) = i \int dkdk' g_{kk'} \left( \hat{a}_{x,k'}^{\dagger} \hat{a}_{y,k} - \hat{a}_{y,k'}^{\dagger} \hat{a}_{x,k} \right) \delta(\eta - c(|k| - |k'|)).
\end{equation}
To calculate the dissipation rate in Eq.~(\ref{def:kappakl}), we need the initial optical density matrix, $\hat{\rho}_{\rm opt}$. 
The optical state consists of a coherent input in $x$-polarization and a vacuum in $y$-polarization, giving
\begin{equation}\begin{aligned}
	\hat{a}_{x,k} \hat{\rho}_{\rm opt} &= \sqrt{2\pi} a_{\rm in} \delta(k-k_{\rm in}) \hat{\rho}_{\rm opt} \\
	\hat{a}_{y,k} \hat{\rho}_{\rm opt} &= 0.
\end{aligned}\end{equation}
The factor of $\sqrt{2\pi}$ makes the definition of the optical amplitude, $a_{\rm in}$, consistent with the analysis in Heisenberg picture above.
Using the commutation relation, $[\hat{a}_{y,k}, \hat{a}_{y,k'}^{\dagger}] = \delta(k-k')$, we get
\begin{equation}
	\kappa_{00} (\eta) = |2\pi a_{\rm in}|^2 \int dk |g_{kk_{\rm in}}|^2 \delta(\eta - c(|k| - k_{\rm in}))
\end{equation}
Looking ahead, typical skyrmion frequency differences are in gigahertz, much smaller than optical frequencies of $\sim \SI{500}{\THz}$. 
Thus, we have $k_{\rm in} + \eta/c \approx k_{\rm in}$, such that we can approximate $g_{kk_{\rm in}} \approx g_{k_{\rm in} k_{\rm in}}$ to get $\kappa_{00}(\eta) = 2\kappa_{\rm opt}$, where $\kappa_{\rm opt}$ was defined above.
The factor of $2$ comes from the two contributions at $k=\pm k_{\rm in}$ modeling decay into forward and backward moving photons, compared to $\kappa_{\rm opt}$ that includes only the photons in transmission.

Inserting these into Eq.~(\ref{res:gen_Gamma}), we find the transition rates as
\begin{equation}
	\Gamma_{uv}^{\rm opt} = 2\kappa_{\rm opt} |\braket{u | \Sz | v}|^2. \label{eq:Gammaopt}
\end{equation}
The off-diagonal decay terms are given by inserting the above into Eq.~(\ref{res:gen_gamma}),
\begin{equation}
    \gamma_{uv}^{\rm opt} = \kappa_{\rm opt} \left( \braket{u|\Sz^2|u} + \braket{v|\Sz^2|v} -2\braket{u|\Sz|u} \braket{v|\Sz|v} \right). \label{eq:gammaopt}
\end{equation}

\subsection{Internal Dissipation} \label{ssec:Int}
Internal dissipation of a skyrmion can be caused by impurity fluctuations and elementary excitations in the surrounding ferromagnetic state. 
For any of these sources, the bath will be in equilibrium, i.e. the the bath's density matrix is $\hat{\rho}_B = e^{-\beta\hat{H}_B}/Z_B$ with the inverse temperature $\beta = 1/k_B T$ and $Z=\mathrm{Tr}[e^{-\beta \hat{H}_B}]$. 
Then, the dissipation constant defined in Eq.~(\ref{def:kappakl}) is,
\begin{equation}
    \kappa_{kl}(\eta) = \frac{2\pi}{Z_B} \int d\gamma d\alpha\ e^{-\beta\hbar\omega_{\gamma}} \delta(\omega_{\beta} - \omega_{\alpha} - \eta) \braket{\gamma | \hat{B}_k|\alpha} \braket{\alpha | \hat{B}_l| \gamma}.
\end{equation}
This formula satisfies a detailed balance condition,
\begin{equation}
    \kappa_{kl}(\eta) = \kappa_{lk}(-\eta) e^{\beta\eta}. \label{cond:bal}
\end{equation}
Using this condition in the master equation, Eq.~(\ref{evol:prob_u}), we get the equation of motion for the diagonal entries of the skrymion's density matrix,
\begin{equation}
	\dot{p}_u = \sum_{kl,v} \kappa_{kl}(\omega_v - \omega_u) \braket{u | \hat{D}_l | v} \braket{v | \hat{D}_k | u} \left( p_v - p_u e^{-\beta(\omega_v - \omega_u)} \right).
\end{equation}
Then, $p_u \propto e^{-\beta \omega_u}$ is a steady state solution, regardless of the choices of $\hat{D}_k$.

Besides these general considerations, the microscopic model of dissipation for skyrmions is dependent on the specific material and sample.
Here, we assume a simple model where baths with different symmetries couple to the lowest order terms, $\Sz {B}_1 + \cos\hat{\varphi} \hat{B}_2 + \sin\hat{\varphi} \hat{B}_3$ where $\hat{B}_i$ are some unknown bath operators. 
We assume them to be independent Ohmic baths of harmonic oscillators, giving $\kappa_{kl}(\eta) = \delta_{kl} \kappa(\eta)$, where
\begin{equation}
    \kappa(\eta) = \alpha_G |\eta|
    \begin{cases}
        n_{\rm BE} (\eta) + 1,\ \eta>0 \\
        n_{\rm BE}(|\eta|),\ \eta<0 
    \end{cases}.
\end{equation}
Here, $\alpha_G$ is the Gilbert damping, and the Bose-Einstein distribution is
\begin{equation}
    n_{\rm BE}(\eta) = \left( e^{\beta \eta} - 1 \right)^{-1},
\end{equation}
with the inverse temperature $\beta$. 
This form ensures the detailed balance condition in Eq.~(\ref{cond:bal}).
With these assumptions, we get
\begin{equation}
    \Gamma_{uv}^{\rm int} =  \kappa(\omega_v - \omega_u)\sum_k \left| \braket{u|\hat{D}_k|v}\right|^2, \label{eq:Gammaint}
\end{equation}
and
\begin{equation}
    \gamma_{uv}^{\rm int} = \frac{1}{2} \sum_e \Gamma_{eu} + \Gamma_{ev} - \frac{\alpha_G}{\beta} \sum_{k} \braket{u | \hat{D}_k | u} \braket{v| \hat{D}_k | v}, \label{eq:gammaint}
\end{equation}
where $\hat{D}_k \in \{\Sz, \cos\Opheli, \sin\Opheli \} $.
Here, we used $\kappa(0) = \alpha_G/\beta$.

\subsection{Spectrum} \label{ssec:Spec}
The optical spectrum is given by a Fourier transform of the correlation function $\langle \Sz(t') \Sz(t) \rangle$. In the steady state, the correlation function is given by the quantum regression theorem~\cite{BP_OQS},
\begin{equation}
    \langle \Sz(\tau) \Sz(0) \rangle = C(\tau) =  \mathrm{Tr}\left[
	\begin{cases}
		\Sz e^{\mathcal{L}|\tau|} \left( \hat{\rho}_e \Sz \right)  & \tau < 0 \\
		\Sz e^{\mathcal{L}\tau} \left( \Sz \hat{\rho}_e \right)  & \tau > 0
	\end{cases} \right].
\end{equation}
The off-diagonal entries decay to zero, and thus the equilibrium state is of the form $\hat{\rho}_e = \sum_u p_u \ket{u}\bra{u}$. 
The populations $p_u$ are calculated by finding the steady-state solution of the master equation, Eq.~(\ref{evol:prob_u}), with $\Gamma_{uv} = \Gamma_{uv}^{\rm int} + \Gamma_{uv}^{\rm opt}$. 
Once the steady-state probabilities are found, we get 
\begin{equation}
    e^{\mathcal{L}\tau} \left( \Sz \hat{\rho}_e \right) = \sum_{uv} p_v \braket{u|\Sz|v} e^{\mathcal{L}\tau}[\ket{u}\bra{v}].
\end{equation}
As argued above, see Eq.~(\ref{evol:off_diag}), $e^{\mathcal{L}\tau}[\ket{u}\bra{v}] = e^{-i(\omega_u - \omega_v)\tau} e^{-\gamma_{uv}\tau} \ket{u}\bra{v}$ for $u\ne v$ where $\gamma_{uv} = \gamma^{\rm opt}_{uv} + \gamma^{\rm int}_{uv}$. 
Thus,
\begin{equation}
    \mathrm{Tr}\left[ \Sz e^{\mathcal{L}\tau} \left( \Sz \hat{\rho}_e \right) \right] = \sum_{u \ne v} p_v \left| \braket{u|\Sz|v} \right|^2 e^{-i(\omega_u - \omega_v)\tau} e^{-\gamma_{uv}\tau} + C_0(\tau),
\end{equation}
where
\begin{equation}
    C_0(\tau) = \sum_u p_u \braket{u|\Sz |u} \mathrm{Tr}\left[\Sz e^{\mathcal{L} \tau}[\ket{u}\bra{u}] \right].
\end{equation}
It is straightforward to calculate $e^{\mathcal{L}\tau}$ in terms of $\Gamma_{uv}$ using the master equation, Eq.~(\ref{evol:prob_u}). 
The term $e^{\mathcal{L}\tau}[\ket{u}\bra{u}]$ will consist of time-dependence of the order of $\Gamma_{uv}$.
For a sufficiently low damping, this time-dependence will be much slower than skyrmion dynamics and therefore, will correspond to a peak near the zero frequency.
This peak will be much smaller than the Faraday rotated light and thus, not possible to resolve in the output photons.

A similar derivation gives
\begin{equation}
    \mathrm{Tr}\left[ \Sz e^{\mathcal{L}\tau} \left( \hat{\rho}_e \Sz \right) \right] = \sum_{u \ne v} p_v \left| \braket{u|\Sz|v} \right|^2 e^{-i(\omega_v - \omega_u)\tau} e^{-\gamma_{vu}\tau} + C_0(\tau)
\end{equation}
Due to Hermiticity of a density matrix, $\gamma_{uv} = \gamma_{vu}$. 
This is also obvious from the explicit expressions of $\gamma^{\rm opt}_{uv}$, Eq.~(\ref{eq:gammaopt}), and $\gamma^{\rm int}_{uv}$, Eq.~(\ref{eq:gammaint}), written above.
Thus, we get
\begin{equation}
    C(\tau) = \sum_{uv} p_v \left| \braket{u|\Sz|v} \right|^2 e^{-i(\omega_u - \omega_v)\tau} e^{-\gamma_{uv}|\tau|} +C_0(\tau).
\end{equation}

The output photons are
\begin{equation}
    \langle \hat{A}_{\omega}^{\dagger} \hat{A}_{\omega} \rangle = \frac{\kappa_{\rm opt}}{T} \int_0^T dt dt' C(t'-t) e^{i\omega(t-t')}, 
\end{equation}
For $T\rightarrow\infty $, we get,
\begin{equation}
    \langle \hat{A}_{\omega}^{\dagger} \hat{A}_{\omega} \rangle = \sum_{uv} \frac{2 \kappa_{\rm opt} p_v \left| \braket{u|\Sz|v} \right|^2\gamma_{uv}}{(\omega - \omega_v + \omega_u)^2 + \gamma_{uv}^2} + N_0(\omega),
\end{equation}
where $N_0(\omega)$ is the Fourier transform of $C_0(\tau)$.

\bibliography{References}

\end{document}